\documentclass[iop]{emulateapj}
\shorttitle{\textsl{Herschel}-ATLAS Galaxy Counts and LFs: Formation of Massive ETGs}
\shortauthors{A. Lapi et al.}

\begin{document}
\title{\textsl{Herschel}-ATLAS Galaxy Counts and High Redshift Luminosity Functions: \\ The Formation of
Massive Early Type Galaxies\footnotemark[$\star$]}\footnotetext[$\star$]{Herschel is an ESA space observatory
with science instruments provided by European-led Principal Investigator
consortia and with important participation from NASA.}
\author{A. Lapi\altaffilmark{1,2},
J. Gonz\'alez-Nuevo\altaffilmark{2}, L. Fan\altaffilmark{2,3}, A.
Bressan\altaffilmark{4,2}, G. De Zotti\altaffilmark{4,2}, L.
Danese\altaffilmark{2}, M. Negrello\altaffilmark{5}, L.
Dunne\altaffilmark{6}, S. Eales\altaffilmark{7}, S. Maddox\altaffilmark{6},
R. Auld\altaffilmark{7},  M. Baes\altaffilmark{8}, D.G.
Bonfield\altaffilmark{9}, S. Buttiglione\altaffilmark{4}, A.
Cava\altaffilmark{10}, D.L. Clements\altaffilmark{11}, A.
Cooray\altaffilmark{12},  A. Dariush\altaffilmark{7,11}, S.
Dye\altaffilmark{7}, J. Fritz\altaffilmark{8}, D. Herranz\altaffilmark{13},
R. Hopwood\altaffilmark{5,7}, E. Ibar\altaffilmark{14},  R.
Ivison\altaffilmark{14,15}, M.J. Jarvis\altaffilmark{9,16},  S.
Kaviraj\altaffilmark{7}, M. L\'opez-Caniego\altaffilmark{13}, M.
Massardi\altaffilmark{17}, M.J. Micha{\l}owski\altaffilmark{15}, E.
Pascale\altaffilmark{7}, M. Pohlen\altaffilmark{7}, E. Rigby\altaffilmark{6},
G. Rodighiero\altaffilmark{18},  S. Serjeant\altaffilmark{5}, D.J.B.
Smith\altaffilmark{9}, P. Temi\altaffilmark{19}, J. Wardlow\altaffilmark{12},
P. van der Werf\altaffilmark{20}} \altaffiltext{1}{Dip. Fisica, Univ. `Tor
Vergata', Via Ricerca Scientifica 1, 00133 Roma, Italy}
\altaffiltext{2}{Astrophysics Sector, SISSA, Via Bonomea 265, 34136 Trieste,
Italy} \altaffiltext{3}{Center for Astrophysics, Univ. of Science and
Technology of China, 230026 Hefei, China} \altaffiltext{4}{INAF-Osservatorio
Astronomico di Padova, Vicolo dell'Osservatorio 5, 35122 Padova, Italy}
\altaffiltext{5}{Dept. of Physics \& Astronomy, The Open Univ., Milton Keynes
MK7 6AA, UK} \altaffiltext{6}{School of Physics and Astronomy, Univ. of
Nottingham, Univ. Park, Nottingham NG7 2RD, UK} \altaffiltext{7}{School of
Physics and Astronomy, Cardiff Univ., The Parade, Cardiff, CF24 3AA, UK}
\altaffiltext{8}{Sterrenkundig Observatorium, Univ. Gent, Krijgslaan 281 S9,
B-9000 Gent, Belgium} \altaffiltext{9}{Centre for Astrophysics, Univ. of
Hertfordshire, Hatfield, Herts, AL10 9AB, UK} \altaffiltext{10}{Dep. de
Astrof\'{\i}sica, Fac. de CC. F\'{\i}sicas, Univ. Complutense de Madrid,
E-28040 Madrid, Spain} \altaffiltext{11}{Astrophysics Group, Imperial
College, Blackett Lab, Prince Consort Road, London SW7 2AZ, UK}
\altaffiltext{12}{Dept. of Physics \& Astronomy, Univ. of California, Irvine,
CA 92697, USA} \altaffiltext{13}{Inst. de Fisica de Cantabria (CSIC-UC),
Avda. los Castros s/n, 39005 Santander, Spain} \altaffiltext{14}{UK Astronomy
Technology Centre, Royal Observatory, Blackford Hill, Edinburgh EH9 3HJ, UK}
\altaffiltext{15}{Inst. for Astronomy, Univ. of Edinburgh, Royal Observatory,
Blackford Hill, Edinburgh EH9 3HJ, UK} \altaffiltext{16}{Physics Dept., Univ.
of the Western Cape, Cape Town, 7535, South Africa}
\altaffiltext{17}{INAF-IRA, Via P. Gobetti 101, I-40129 Bologna, Italy}
\altaffiltext{18}{Dip. Astronomia, Univ. di Padova, Vicolo dell'Osservatorio
3, I-35122 Padova, Italy}\altaffiltext{19}{Astrophysics Branch, NASA Ames
Research Center, Mail Stop 245-6, Moffett Field, CA 94035, USA}
\altaffiltext{20}{Sterrewacht Leiden, Leiden Univ., PO Box 9513, 2300 RA
Leiden, The Netherlands}

\begin{abstract}
Exploiting the \textsl{H-ATLAS} Science Demonstration Phase (SDP) survey
data, we have determined the luminosity functions (LFs) at rest-frame
wavelengths of $100$ and $250\, \mu$m and at several redshifts $z\ga 1$, for
bright sub-mm galaxies with star formation rates (SFR) $\ga 100\,
M_{\odot}\,\mathrm{yr}^{-1}$. We find that the evolution of the comoving LF
is strong up to $z\approx 2.5$, and slows down at higher redshifts. From the
LFs and the information on halo masses inferred from clustering analysis, we
derived an average relation between SFR and halo mass (and its scatter). We
also infer that the timescale of the main episode of dust-enshrouded star
formation in massive halos ($M_H\ga 3\times 10^{12}\, M_{\odot}$) amounts to
$\sim 7\times 10^8$ yr. Given the SFRs, which are in the range $10^{2}-10^3\,
M_{\odot}\,\mathrm{yr}^{-1}$, this timescale implies final stellar masses of
order of  $10^{11}-10^{12}\, M_{\odot}$. The corresponding stellar mass
function matches the observed mass function of passively evolving galaxies at
$z\ga 1$. The comparison of the statistics for sub-mm and UV selected
galaxies suggests that the dust-free, UV bright phase, is $\ga 10^2$ times
shorter than the sub-mm bright phase, implying that the dust must form soon
after the onset of star formation. Using a single reference Spectral Energy
Distribution (SED; the one of the $z\approx 2.3$ galaxy SMM J2135-0102), our
simple physical model is able to reproduce not only the LFs at different
redshifts $> 1$ but also the counts at wavelengths ranging from $250\,\mu$m
to $\approx 1$ mm. Owing to the steepness of the counts and their relatively
broad frequency range, this result suggests that the dispersion of sub-mm
SEDs of $z>1$ galaxies around the reference one is rather small.
\end{abstract}


\section{Introduction}\label{sect:intro}

The star formation history in galaxies is one of the key issues we have to
understand in order to reconstruct how the Universe evolved from small matter
perturbations at the recombination epoch to the present richness of
structures.

Star formation proceeds at a different pace for different galaxies, depending
on the physical conditions of the available gas. Early Type Galaxies (ETGs)
and massive bulges of S$a$ galaxies are composed of relatively old stellar
populations with mass-weighted ages of $\ga 8-9$ Gyr (corresponding to
formation redshifts $z\ga 1-1.5$), while the disc components of spiral and
irregular galaxies are characterized by significantly younger stellar
populations. For instance, the luminosity-weighted age for most of S$b$ or
later-type spirals is $\la 7$ Gyr (cf. Bernardi et al. 2010, their Fig.~10),
corresponding to a formation redshift $z\la 1$. In general, the old stellar
populations feature low specific angular momentum as opposed to the larger
specific angular momentum of the younger ones.

How can these facts be interpreted in the framework of the hierarchical
evolution of the dark matter (DM) galaxy halos that have proven to be
remarkably successful in accounting for the power-spectrum (or the spatial
correlation function) of the large-scale matter distribution (e.g. Springel
et al. 2006)? A widely held view is that the merging of halos is also the
principal mechanism driving the evolution of the visible part of galaxies
(see Benson 2010 for a recent review). However, several bodies of evidence
are difficult to reconcile with this scenario.

\begin{itemize}

\item ETGs are characterized by old and homogeneous stellar populations.
    Correlations tight enough to allow little room for random processes
    such as a sequence of mergers (apart from small mass additions
    through minor mergers at late epochs, see Kaviraj et al. 2008), and
    sensitivity to the environment (color$-$luminosity; fundamental plane
    relations; dynamical mass$-$luminosity) have been known for a long
    time and have been recently confirmed with very large samples, and
    shown to persist up to substantial redshifts (Renzini 2006; Clemens
    et al. 2009; Thomas et al. 2010; Rogers et al. 2010; Peebles \&
    Nusser 2010, and references therein). More recently, a remarkably
    tight luminosity$-$size correlation has been reported (Nair et al.
    2010). In addition, ETGs were found to host supermassive black holes
    whose mass is proportional to the bulge and to the halo mass of the
    host galaxy (see Magorrian et al. 1998; also Ferrarese \& Ford 2005
    for a review). All that indicates that the formation and evolution of
    ETGs is almost independent of environment, and driven mainly by
    self-regulation processes and intrinsic galaxy properties such as
    mass.

\item There are rather tight, albeit not inescapable, observational
    constraints on the star-formation timescale in the most massive ETGs.
    An upper limit comes from the observed $\alpha$-enhancement or, more
    properly, iron under-abundance compared to $\alpha$ elements.
    Depending on the slope of the assumed initial mass function (IMF),
    the observed $\alpha/${\it Fe} element ratios requires star formation
    timescales $\la 10^9$ yr (e.g. Matteucci 1994; Thomas et al. 1999).
    But in merger-driven galaxy formation  models star formation in
    ellipticals typically does not truncate after 1 Gyr (Thomas \&
    Kauffmann 1999; however, see Arrigoni et al. 2010, Khochfar \& Silk
    2010). \emph{If a standard IMF is assumed} a lower limit comes from
    (sub-)mm counts, implying that several percent of massive galaxies
    are forming stars at rates of thousands $M_\odot$ yr$^{-1}$ at $z\sim
    2-3$ (see Chapman et al. 2003, 2005). This requires that this
    star-formation rate (SFR) is sustained for $\ga 0.5$ Gyr, much longer
    than the timescale of a merger-induced starburst, which is of order
    of the dynamical time ($\sim 0.1$ Gyr for the massive ETGs of
    interest here; see, e.g., Benson 2010; Hopkins 2011). In other words,
    a single starburst episode is too short to account for the space
    density of (sub-)mm bright galaxies as well as for their present-day
    stellar masses. And indeed models envisage a sequence of mergers, and
    associated starbursts, throughout the galaxy lifetime, i.e. over a
    time span much longer than the upper limit set by the
    $\alpha$-enhancement. The problem of accounting for the counts of
    sub-mm galaxies may be eased assuming a top-heavy IMF (Baugh et al.
    2005). Indeed, recent observational evidences (Gunawardhana et al.
    2011; Dunne et al. 2011) indicate that highly star forming galaxies
    have IMFs $dN/dm \propto m^{-x}$ with flatter high-mass power-law
    slopes $x$ than galaxies with low star formation rates. Gunawardhana
    et al. (2011), using a sample of galaxies from the Galaxy And Mass
    Assembly (GAMA) survey (Driver et al. 2009) covering the redshift
    range $0<z<0.35$, find a dependence of $x$ on the SFR that,
    extrapolated to an $SFR=1000\, M_\odot$ yr$^{-1}$, would give
    $x\approx 1.5$, substantially flatter than the Salpeter (1955) slope
    ($x_S=2.35$). However, it is not clear that these results apply to
    the high-$z$ proto-spheroidal galaxies considered in this paper,
    since the IMF may depend on other parameters, such as age and
    metallicity. Moreover, the most recent study of the evolution of
    galaxies in the far-infrared/sub-mm based on starbursts triggered by
    mergers (Lacey et al. 2010) resorts to an even flatter high-mass IMF
    ($x=1$) and has still serious problems with reproducing the
    \textsl{Herschel} counts (see \S\,\ref{sect:counts}).

\item Integral-field near-IR spectroscopy of galaxies with less extreme
    SFRs (of few $10^2\,M_\odot$ yr$^{-1}$) at $z\sim 2$, that appear to
    be very productive star formers in the universe (Dekel et al. 2009),
    has shown that in many cases they  have ordered, rotating velocity
    fields with no kinematic evidence for ongoing merging (Genzel et al.
    2006; F\"orster-Schreiber et al. 2009). Still, they harbor several
    starforming clumps: a complex morphology is not necessarily a symptom
    of merging. These galaxies show tight SFR$-$mass correlations, with
    small dispersions (Daddi et al. 2007; Pannella et al. 2009; Dunne et
    al. 2009; Rodighiero et al. 2010; Maraston et al. 2010). This is not
    easily reconciled with a scenario in which star formation proceeds
    through a series of short starbursts interleaved by long periods of
    reduced activity and these galaxies have been caught in a special,
    starburst moment of their existence. The data are more easily
    accounted for if the high SFRs are sustained for some $1-2$ Gyr, much
    longer that a dynamical time typical of starbursts. Although
    the duration of the star-formation phase for these objects is longer
    than that of the more extreme objects considered in the previous
    bullet by a factor of 2 to 3, their final stellar mass is several
    times lower because the SFR is about an order of magnitude smaller.
    For these galaxies a weak $\alpha$-enhancement is predicted,
    consistent with observations.

\item A comparison of the stellar mass functions at $z\ga 1.5$ with the
    local one shows that little additional growth can have occurred for
    $z \la 1.5$ through minor mergers (Mancone et al. 2010; Fan et al.
    2010; Kaviraj et al. 2009).

\end{itemize}

It is clear from the above that the reconstruction of the star formation
history through cosmic time for galaxies of different masses provides a
critical test for galaxy formation and evolution. Since the star formation
occurs within dusty molecular clouds and is deeply obscured at ultraviolet
and optical wavelengths, data at far-IR/(sub-)mm wavelengths, where the
absorbed radiation is re-emitted, are essential to provide a complete
picture of it.

In this paper we focus on high redshift ($z\ga 1$) galaxies. We exploit the
far-IR/sub-mm data collected by the \textsl{Herschel} Space Observatory
(Pilbratt et al. 2010) during the Science Demonstration Phase (SDP) of the
\textsl{Herschel} Astrophysical Terahertz Large Area Survey
(\textsl{H-ATLAS}, Eales et al. 2010a). The \textsl{H-ATLAS} is an open-time
key program that will survey $\sim 550\,\mathrm{deg}^2$ with PACS (Poglitsch
et al. 2010) and SPIRE (Griffin et al. 2010) in five bands, from $100$ to
$500\,\mu$m.

The observed SDP field covers an area of $\approx 3.8\times
3.8\,\mathrm{deg}^2$ centered on $(\alpha,\delta)\approx $ $(09^{\rm h}\,
05^{\rm m},$ $+0^\circ\, 30')$. Complete descriptions on reduction of PACS
and SPIRE data are given in Ibar et al. (2010) and Pascale et al. (2011),
respectively. Source extraction and flux density estimation are described in
Rigby et al. (2011). The $5\sigma$ detection limits,  including confusion
noise, are $33.5$, 37.7, and 44.0 mJy/beam in the SPIRE bands at 250, 350,
and $500\,\mu$m, respectively; in the PACS bands they are 132 mJy and 121 mJy
at 100 and $160\,\mu$m, respectively.

The plan of the paper is the following. The selection of the sample is
described in \S~\ref{sect:sample}. In \S~\ref{sect:sed} we discuss the far-IR
Spectral Energy Distribution (SED) of high-$z$ star forming galaxies, a
fundamental ingredient for our photometric redshift estimates, presented in
\S~\ref{sect:redshift}. In \S~\ref{sect:LF} we estimate the galaxy luminosity
functions (LFs) at different redshift in the range $1-4$. In
\S~\ref{sect:disc} we discuss some clues on star-formation timescales for
massive galaxies. In \S~\ref{sect:counts} model predictions for source counts
from $250\,\mu$m to 2 mm are compared with the data.  Our main conclusions
are summarized in \S~\ref{sect:concl}.

Throughout the work we adopt a standard, flat $\Lambda$CDM cosmology (see
Komatsu et al. 2011) with matter density parameter $\Omega_M=0.27$ and Hubble
constant $H_0=70\,\mathrm{km}\,\mathrm{s}^{-1}\,\mathrm{Mpc}^{-1}$. We adopt
a Bardeen et al. (1986) cold dark matter power spectrum with primordial index
$n_s=1$ and cosmic mass variance $\sigma_8=0.81$. Stellar masses and
luminosities of galaxies are evaluated assuming the Chabrier's (2003) initial
mass function (IMF); these can be converted to a Salpeter (1955) IMF on
multiplying by a factor $\approx 1.6$.

\setcounter{footnote}{0}

\section{Sample selection}\label{sect:sample}

The \textsl{H-ATLAS} sources comprise both a low-$z$ galaxy population,
identified through matching to the Sloan Digital Sky Survey (SDSS $–$ York et
al. 2000) data (Smith et al. 2011a), and a high-$z$ population (median
redshift $\sim 2$) identified through their far-IR colours (Amblard et al.
2010). Low-$z$ galaxies are generally normal/star-forming late-type galaxies
with moderate opacity (Smith et al. 2011b; Dunne et al. 2011). Through
analyses of clustering these two populations are found to be very different;
the low-$z$ population ($z<0.3$) clusters like star-forming blue galaxies
(Guo et al. 2011; van Kampen et al. 2011; Maddox et al. 2010), while the
high-$z$ population clusters much more strongly, suggesting that the high-$z$
sources reside in more massive halos (Maddox et al. 2010).

In this work we investigate the evolution with cosmic time of high redshift
galaxies with intense star-formation activity, interpreted as massive
proto-spheroidal galaxies in the process of forming most of their stellar
mass (Granato et al. 2004; Lapi et al. 2006). Since these objects are
observed to be in passive evolution at $z\la 1$--1.5, we confine ourselves to
$z>1$. At these redshifts sources above the \textsl{H-ATLAS} detection limits
have dust luminosities $\ga 10^{12}\,L_\odot$ and star-formation rates (SFRs)
$\ga 100\,M_\odot$ yr$^{-1}$.  They are therefore ultra-luminous infrared
galaxies (ULIRGs). Their dust heating mostly comes from young massive stars
within molecular clouds, implying, on one side, that their far-IR SEDs are
generally (albeit not always, e.g. Hwang et al. 2010) warmer than those of
low-$z$ dusty galaxies which have higher contributions from cooler
interstellar dust heated by old stars, and, on the other side, that their
optical emission is strongly attenuated. The last point means not only that
spectroscopic redshifts are available just for a tiny fraction of sources,
but also that we do not have at our disposal the multi-frequency
optical/near-IR photometry that allowed photometric redshift estimates at
lower $z$ (Dye et al. 2010; Vaccari et al. 2010; Eales et al. 2010b).

In fact, for most \textsl{H-ATLAS} $z\ga 1$ galaxies, the only available data
is the \textsl{Herschel} photometry, primarily in the three SPIRE bands
($250$, $350$, and $500\,\mu$m), plus mostly upper limits in the PACS $100$
and $160\,\mu$m bands. A key issue is then whether the redshift estimates
that can be obtained from such data are sufficient to obtain meaningful
estimates of the LFs at least over a limited redshift range. At first sight
one would be inclined to answer `no', but a closer investigation can suggest
a more optimistic conclusion.

The dust re-radiation in starburst galaxies is expected to come from at least
three astrophysical settings (e.g. Silva et al. 1998): molecular clouds,
diffuse low-density clouds (cirrus), and circum-nuclear regions, heated by
Active Galactic Nuclei (AGNs). The AGN dust emission peaks in the mid-IR
(Granato \& Danese 1994; Andreani et al. 2010; Lutz et al. 2010;
Hatziminaoglou et al. 2010) and can be safely ignored in the SPIRE wavelength
range. Molecular clouds are the preferential sites of star formation implying
that they are endowed with intense radiation fields and relatively warm dust
temperatures. The cirrus component is exposed to the less intense general
radiation field due to older stellar populations that have come out from
their native molecular clouds and is therefore characterized by lower dust
temperatures.

In the nearby Universe, molecular clouds and cirrus give comparable
contributions to the far-IR emission from `normal' late-type galaxies, with
relatively low SFRs (Rowan-Robinson et al. 2005). The colder cirrus
contribution is especially important in less optically obscured IR galaxies
(Hwang et al. 2010) while the warmer molecular cloud (sometimes referred to
as `starburst') component becomes increasingly important for higher and
higher SFRs (Rowan-Robinson et al. 2010). This argument also highlights a
possible degeneracy: a `cold' observed SED may be associated either to a
low-$z$ cirrus dominated galaxy or to a redshifted warm galaxy. If the
redshift is estimated using a warm SED, cold low-$z$ galaxies would be
erroneously assigned high redshifts. This problem can be overcome, however,
because cold, low-$z$ galaxies are only moderately obscured by dust (the
cirrus optical depth cannot be very large), and are therefore relatively
bright in the optical bands. This is illustrated by Fig.~\ref{fig:rmag},
which shows that the SEDs of optically identified $z<0.5$ SDP galaxies
studied by Smith et al. (2011b) imply $r$-band magnitudes brighter than the
SDSS DR7 limit ($r=22.4$) at all redshifts even if their $250\,\mu$m flux
density is at the detection limit, while $z>1$ ULIRGs, with SED like that of
Arp220, (and even more, younger high-$z$ galaxies with SED like that of SMM
J2135-0102, `The Cosmic Eyelash'; Ivison et al. 2010a, Swinbank et al. 2010)
are fainter than this magnitude limit for the same $250\,\mu$m flux
density\footnote{Note however that a galaxy with the real Arp220 luminosity
would be brighter than $r=22.4$ up to $z\approx 0.6$. At $z=0.5$ it would
have $r\approx 21.5\,$mag. At the same redshift, the true, non-demagnified
SMM J2135-0102, would have $r\approx 19.6\,$mag. On the other hand, high-$z$
moderately obscured galaxies can be detected by the \textsl{H-ATLAS} survey
only if they have very high stellar masses.}.

Therefore, we may weed out \emph{cold} low-$z$ galaxies by dropping SDP
galaxies with SDSS counterparts (Smith et al. 2011a), except those with
optical (spectroscopic or photometric) redshifts in the range of interest
here ($z>1$). This operation, however, has potential drawbacks. First, it
leaves in low-$z$ ULIRGs. This is not a big problem, since these objects have
warm SEDs and therefore, as discussed below, their redshifts can be estimated
with sufficient accuracy. Second, the reliability of SDSS counterparts can
never be $100\%$. For example, strongly lensed galaxies generally have an
apparently reliable counterpart which is most likely the foreground lens.
These objects can however be recovered since most frequently the lenses are
ellipticals, whose optical colours are incompatible with a large dust
emission. The fraction of false identifications among the optical
counterparts to \textsl{H-ATLAS} SDP sources with reliability $R > 0.8$ is
estimated to be $\approx 5.8\%$ (Smith et al. 2011a). Although this fraction
is reassuringly small, we need to keep in mind that some truly high-$z$
sources can be missed by our procedure.

Most importantly, not all the true $r<22.4$ counterparts can be
identified with $>80\%$ confidence, due to incompleteness of the SDSS
catalog, positional uncertainties, close secondaries, and the random
probability of finding a background source within that search radius (Dunne
et al. 2011). According to Smith et al. (2011a), about 60\% of the 6621 with
$250\,\mu$m flux density $> 32\,$mJy have counterparts brighter than $r =
22.4\,$mag in the SDSS (and are therefore, with few exceptions, at $z<1$). Of
these, 2423 could be identified with a reliability $R > 0.8$, implying that
another $\approx 1550$ sources, i.e. about 23\% of the total sample are
really at $z<1$ but are missing a reliable identification. From the redshift
distribution of reliable identifications (Fig. 6 of Smith et al. 2011a) we
estimate that about 20\% of these sources are at $z>0.5$. If the same
proportion applies to unidentified sources, the identification incompleteness
of  $z<0.5$ \textsl{H-ATLAS} SDP galaxies is $\approx 19\%$, that we
conservatively round to 20\%. The effect of this incompleteness on our
luminosity function estimates is discussed in \S\,\ref{sect:LF}.

Starting from the catalog of SDP sources by Rigby et al. (2011), which
contains 6876 sources, we drop the galaxies for which Smith et al. (2011a)
have identified reliable counterparts. We further require detection at $\ge
3\sigma$ at $350\,\mu$m. In this way we get a sample (that will be taken as
our reference sample) defined by the following criteria: i) $S_{250\mu\rm
m}\ge 35\,$mJy; ii) no optical identification with $R>0.8$ (Smith et al.
2011a); iii) detection at $\ge 3\sigma$ at $350\,\mu$m. The resulting sample
is made of $3469$ sources. The redshift estimates presented in
\S\,\ref{sect:redshift} indicate that some of these ($376$) are $z<1$ ULIRGS
with SEDs akin to our templates; they will be excluded from the subsequent
analysis.

Most ($2763$, i.e., $\approx 80\%$) galaxies satisfying the above criteria
are detected at $\ge 4\sigma$ at $350\,\mu$m, and this obviously helps with
the photometric redshift estimates. This is because low-$z$ galaxies have low
$S_{350\mu\rm m}/S_{250\mu\rm m}$ ratios and are therefore lost when we raise
the $350\,\mu$m flux limit (see also Amblard et al. 2010).

\section{SEDs of high-$z$ sub-mm galaxies}\label{sect:sed}

For the objects of interest here, with $\mathrm{SFR}\ga 100\, M_\odot$
yr$^{-1}$, only the `warm' (starburst) component is relevant. It is important
to take into account, however, that its SED is much broader than a
single temperature grey-body (see, e.g., Silva et al. 1998). Over the
$50-500\,\mu$m range (in the rest-frame) such a SED can be approximated, to
better than $10-20\%$, by a sum of two grey-bodies, each described by
\begin{equation}\label{eq:sdust}
S_{\nu}\propto \frac {\nu ^{3+\beta}}{e^ {h\nu/kT_d}-1}~,
\end{equation}
with temperatures $T_d\approx 30$ K and $\approx 60$ K, and dust emissivity
indices $\beta=1.7$ and 2, respectively (see also Dunne \& Eales 2001). A fit
to the observed SEDs of standard template starburst galaxies with quite
different SFRs (M82, Arp220), and of the $z\approx 2.3$ strongly lensed
galaxy SMM J2135-0102, confirms the validity of this approximation and shows
that the relative normalization of the two grey bodies varies by factors of
several. Mid/far-IR data on starburst galaxies emphasize the warmer
component, while sub-mm data emphasize the colder one; if a single
temperature dust is used, data in different wavelength ranges may yield
substantially different temperatures. If the source redshift is in the range
$1\la z\la 3.5$, PACS and SPIRE wavelengths extend across the dust emission
peak (which is typically at a rest-frame wavelength  $\lambda\approx 90-100\,
\mu$m). This may allow reasonably accurate redshift estimates using only
\textsl{Herschel} data (Negrello et al. 2010). However, only a minor fraction
of the \textsl{H-ATLAS} SDP galaxies have at least one PACS detection,
usually at $160\, \mu$m, and these are mostly at low redshifts; we must also
beware of the flux boosting by confusion, increasing at longer wavelengths
(see \S\,\ref{sect:redshift}).

Arp220 and SMM J2135-0102 are of particular interest because their SEDs are
well determined and have SFRs quite typical of the galaxies considered here.
We have modeled their spectral energy distributions from extreme-UV to radio
frequencies through the spectrophotometric code GRASIL\footnote{For
information on the code see
\texttt{http://adlibitum.oat.ts.astro.it/silva/default.html}, and for a
web-based version see \texttt{http://galsynth.oapd.inaf.it}}, which includes
a sophisticated treatment of dust reprocessing (Silva et al. 1998; Schurer et
al. 2009; Silva et al. 2011). The results are illustrated in
Fig.~\ref{fig:SED}; it must be stressed that the dominant contribution for
wavelengths $\lambda\ga 30\, \mu$m is provided by molecular clouds (more
details in Bressan et al. 2011, in prep.). Assuming a Chabrier (2003) IMF we
obtain a SFR of $\approx 140\,M_{\odot}\,\mathrm{yr}^{-1}$ for Arp220 and of
$\approx 270\,M_{\odot}\,\mathrm{yr}^{-1}$ for SMM J2135-0102 (a Salpeter IMF
would yield $\approx 224\,M_{\odot}\,\mathrm{yr}^{-1}$ and $\approx
430\,M_{\odot}\,\mathrm{yr}^{-1}$, respectively). Then assuming the usual
linear scaling between the SFR and the continuum far-IR luminosity $L_{\rm
FIR}$ integrated between $8\,\mu$m and $1000\, \mu$m, i.e.,
\begin{equation}\label{eq:conv}
{\mathrm{SFR}\over M_{\odot}\,{\rm yr}^{-1}}=k\times 10^{-44}\,{L_{\rm
FIR}\over {\rm erg\, s}^{-1}}~,
\end{equation}
we find $k_{\rm Arp}=2.5$ and $k_{\rm SMM}=3.4$. The difference between the
two coefficients is well within the expected range. Kennicutt (1998) pointed
out that variations of $\approx 30\%$ can be due to differences in the
star-formation history, implying different effective ages of the stellar
populations. In Fig.~\ref{fig:SED} we also sketch the far-IR SED of G15.141
(H-ATLAS J142413.9+022304), a strongly lensed sub-mm galaxy at $z\approx
4.24$ with estimated SFR of several hundreds $M_{\odot}$ yr$^{-1}$ (Cox et
al. 2011), modeled as the sum of two grey-bodies with $T_1=32\,$K,
$T_2=60\,$K, $\beta=2$, and a ratio of 0.02 between the coefficients of the
warm and of the cold components. We choose the SEDs of these 3 galaxies
(Arp220, SMM J2135-01012, and G15.141) as our references to quantify the
effect of different choices for the SED on estimates of the LFs. SMM
J2135-01012 has the smallest $S_{60\mu\rm m}/S_{100\mu\rm m}$ ratio and
G15.141 exhibits the steepest decrease at long wavelengths.

\section{Estimating the redshifts of sub-mm galaxies}\label{sect:redshift}

A major source of concern is the boosting of fluxes in the Rigby et al.
(2011) catalogue due to confusion by faint sources. Since the effect is
larger at longer wavelengths, because of the poorer angular resolution, it
tends to bias high the redshift estimates based on \textsl{Herschel}
photometry. According to the simulations carried out by Rigby et al. (2011),
$56.5\%$ of sources detected at $\ge 5\sigma$ at $500\mu$m show a flux
boosting by a factor $>1.5$ and $27.3\%$ by a factor $>2$, and the effect
becomes negligible for $\ge 10\sigma$ sources.

To investigate the quantitative impact of the flux boosting we estimated the
photometric redshifts of 39 \textsl{H-ATLAS} galaxies at $z> 0.5$ for which
spectroscopic redshifts are available. The results are presented in
Fig.~\ref{fig:zspec} for our 3 reference SEDs. The average SED of low-$z$
\textsl{H-ATLAS} SDP galaxies, determined by Smith et al. (2011b), was also
used, for comparison. There is no indication that photometric redshifts of
the high-$z$ sources are systematically overestimated when we use the SED of
SMM J2135-0102 as a template. The median value of $\Delta z/(1+z)\equiv
(z_{\rm phot}-z_{\rm spec})/(1+z_{\rm spec})$ is $0.01$ with a dispersion of
0.21. The situation is only slightly worse in the case of Arp220: the median
value of $\Delta z/(1+z)$ is 0.06 with a dispersion of 0.27. The median
offset between photometric and spectroscopic redshifts increases to 0.18,
with a dispersion of 0.28, if we use the cooler SED of G15.141. In the
redshift range ($z\ge 1$) we are most interested in there are 24 objects. The
median values of $\Delta z/(1+z)$ are $-0.002$ (dispersion 0.12) for SMM
J2135-0102, 0.07 (dispersion 0.16) for Arp220, 0.16 (dispersion 0.13) for
G15.141. There is no statistically significant difference between the 6
strongly lensed objects, having $>10\sigma$ detections at $500\mu$m, and the
other 33 sources, whose flux densities are representative of those in our
sample. The median values of $\Delta z/(1+z)$ and the dispersions rapidly
increase as we go down in redshift, as expected since our templates do not
match those of low-$z$ galaxies whose SEDs include large cirrus
contributions. For the $1096$ \textsl{H-ATLAS} SDP sources with $z<0.5$ and
spectroscopic redshift in the Smith et al. (2011a) catalog we find median
$\Delta z/(1+z)$ of 0.38 (dispersion 0.45), 0.42 (dispersion 0.54), and 0.63
(dispersion 0.55) for SMM J2135-0102, Arp220, and G15.141, respectively. Not
surprisingly, the redshift estimates based on the mean low-$z$ SED of Smith
et al. (2011a) go in the opposite direction: the redshifts are systematically
underestimated. The mean offset is small  ($\Delta z/(1+z)=-0.08$) at $z<0.5$
and increases (in absolute value) to $-0.21 $ for $z>0.5$, and to $-0.34 $
for $z>1$. The dispersions are 0.31, 0.29, and 0.23 for $z<0.5$, $z>0.5$, and
$z>1$, respectively. This confirms that the low-$z$, optically bright,
\textsl{H-ATLAS} galaxy population has far-IR properties different from those
of high-$z$ galaxies. Hence, the SEDs determined at low-$z$ are not
applicable at high-$z$.

This test suggests that, at least for sources at $z\ga 1$, errors on
photometric redshift estimates are more related to the choice the SED
template than to the signal-to-noise ratio at $500\mu$m, and hence to flux
boosting. This is mostly due to the fact that the relative errors on flux
densities are larger at longer wavelengths so that the data points more
liable to flux boosting weight less in the minimum $\chi^2$ fit to the
template SED. We do see in most cases that the best fit SED is below the
$500\mu$m data, as expected if the latter is overestimated. We have checked
this by lowering the $500\mu$m flux densities first by $20\%$, and this left
the derived redshift distribution almost unchanged, and then by a factor of
2. In the latter case, we got a median value of $(z_{\rm
deboost}-z)/(1+z)\approx -0.1$, with only $5\%$ of cases at $<-0.2$ (here $z$
is the redshift estimated using the fluxes tabulated by Rigby et al. 2011).

A second concern is the effect of the SED variety of active star forming
galaxies. To quantify the corresponding errors in the redshift estimates, we
generated a catalogue of $9\times 10^3$ galaxies with a distribution of flux
densities reflecting that of our sample and redshifts randomly selected in
the range $1\la z\la 3.5$. Each redshift was randomly assigned to one of the
$19$ well observationally determined SEDs of local (ultra-)luminous IR
galaxies with star formation rates $\ga 20\, M_{\odot}$ yr$^{-1}$ and with
contributions from an active nucleus to the far-IR flux $f_{\rm AGN}\la 10\%$
studied by Vega et al. (2008). To the $250$, $350$, and $500\, \mu$m flux
densities we associated errors extracted randomly from the distribution of
the errors for real observations. We have then estimated the redshifts of
simulated galaxies, based on flux densities at SPIRE wavelengths, using each
of our three template SEDs.

In $\approx 95\%$ of the cases we have $|z_{\rm phot}-z_{\rm true}|/(1+z)\la
0.2$. Note that this is likely an upper limit since the previous test and the
analysis of multi-frequency source counts (see \S\,\ref{sect:counts})
strongly indicate that the SEDs of high-$z$ galaxies with intense
star-formation are more uniform, and closer to that of SMM J2135-01012, than
those of star-forming low-$z$ galaxies. This can be expected since these
galaxies have generally lower star-formation rates (of order of tens
$M_\odot/$yr, to be compared with $> 100\,M_\odot$ yr$^{-1}$ for high-$z$
galaxies) and higher contributions to dust heating from old stellar
populations. As the SFR increases, the ratio between the contributions
to the far-IR/SMM SED from the molecular cloud component (associated to
star-formation) and from the cirrus component (heated by the general
interstellar radiation field) increases causing a systematic variation of the
SED shape with the specific star formation rate, as indeed observed (da Cunha
et al. 2008; Smith et al. 2011b). This trend stops when the former component
dominates, as in the case of sources considered in this paper, and the SEDs
become more uniform.

We have estimated the redshifts\footnote{The estimated redshifts are
available at http://people.sissa.it/$\sim$gnuevo/photoz/.} of objects in the
samples defined in \S\,\ref{sect:sample}. Sources undetected at $\ga 3\,
\sigma$ at $500\,\mu$m were attributed a $3\sigma$ upper limit of $27$ mJy.
The calculations were repeated using $5\sigma$ upper limits (45 mJy): the
derived redshift distribution did not change appreciably. The results are
similar, but somewhat more sensitive to the effect of boosting (based on the
tests described above), if we use the catalogued flux densities and errors
also for sources with  $S_{500\,\mu\rm m} < 3\sigma$. The redshift estimate
is the result of a minimum $\chi^2$-fit of each of the three templates to the
SPIRE and PACS data (including $3\sigma$ upper limits\footnote{The results,
however, are only weakly constrained by PACS data. We do not find significant
differences if we use $5\sigma$ upper limits or ignore these data
altogether.}).  Fig.~\ref{fig:zdist} shows that, after correcting for the
offsets highlighted by Fig.~\ref{fig:zspec}, the derived redshift
distributions are only moderately affected by the choice of the template SED;
they have broad maxima in the range $1.5\la z \la 2.5$ and a tail  extending
up to $z\approx 3.5$, consistent with earlier estimates for BLAST (Ivison et
al. 2010b) and \textsl{Herschel} (Amblard et al. 2010; Eales et al. 2010)
samples, when the different selections, and in particular the fact that we
have dropped galaxies with $R>0.8$ optical counterparts, are taken into
account. Sources with estimated $z<1$ will be excluded from the subsequent
analysis. With our preferred SED, that of SMM J2135-01012,  our reference
sample ($S_{250\mu\rm m}>35\,$mJy, $S_{350\mu\rm m}>3\sigma$, and no $R>0.8$
optical identifications) contains 3093 galaxies (i.e. $\approx 45\%$ of the
6876 \textsl{H-ATLAS} SDP  galaxies) at $z>1$, consistent with the
fraction of sources expected to be below the SDSS limit ($\sim 40\%$) and
therefore at high $z$. Only 33 of them have $\ge 5\sigma$ detections in at
least one PACS channel; these include the strongly lensed galaxies found by
Negrello et al. (2010).

\section{Luminosity function of high-$z$ sub-mm galaxies}\label{sect:LF}

We have computed the LFs at rest-frame wavelengths of $100$ and $250\,\mu$m
in $4$ redshift intervals ($1.2\le z < 1.6$, $1.6\le z < 2$, $2\le z < 2.4$,
$2.4\le z < 4$), exploiting the classical Schmidt's (1968) $1/V_{\rm max}$
estimator, our redshift estimates, and $K-$corrections computed from the SEDs
used for the redshift estimates. The $V_{\rm max}$ was computed, for each
galaxy, taking into account the redshift boundaries of the bin and the
maximum accessible redshifts implied by the 250 and $350\,\mu$m flux density
limits. In the case of our preferred SED (SMM J2135-01012) the numbers of
sources in these redshift bins are 900, 891, 821, and 502, respectively.

The error on $z$ introduces both a statistical and a systematic effect. The
latter is related to the Eddington bias, that can be quantified with a
Bayesian approach. The probability that the true redshift of a source is $z$
when its estimated value is $z_e$ reads
\begin{equation}
p(z|z_e) \propto P(z_e|z)\,n(z)
\end{equation}
where $P(z_e|z)$ is the probability that the estimated redshift of a source
is $z_e$ when its true value is $z$ and $n(z)$ is the true redshift
distribution. We take $P(z_e|z)$ to be a Gaussian with mean $z$ and
dispersion equal to the rms difference between spectroscopic and photometric
redshifts for the galaxies in Fig.~\ref{fig:zspec} and $n(z_e)$ as a proxy
for $n(z)$. The maximum likelihood estimate of $z$ is the value for which
$p(z|z_e)$ peaks. All the luminosity functions presented in this paper are
corrected for this bias. {We caution, however, that this correction does not
entirely account for the Eddington bias due to the errors on estimated
luminosities. The effect is expected to be smaller than that of the error on
$z$, which is the main source of the error on luminosity. A full treatment
would require extensive numerical simulations that are beyond the scope of
this paper.

As mentioned in \S\,\ref{sect:sample}, we expect a residual contamination of
our reference sample by cold low-$z$ galaxies with genuine $r<22.4$
counterparts that could not be identified with $R>0.8$. To quantify the
effect of this contamination on our luminosity function estimates we have
exploited the \textsl{H-ATLAS} SDP sources with spectroscopic redshift in the
Smith et al. (2011a) catalog. Out of a total of $1096$ galaxies with $z_{\rm
spec}<0.5$, 682 comply with our selection criteria ($S_{250\mu\rm m}>35\,$mJy
and $S_{350\mu\rm m}>3\sigma$).  For these sources the use of the SMM
J2135-0102 template for redshift estimates is inappropriate and yields a
substantial positive offset and a rather broad dispersion of $z_{\rm
phot}-z_{\rm spec}$ (see \S\,\ref{sect:redshift}). In fact, after the
Bayesian correction described above, this template yields $z_{\rm phot}\ge
1.2$ for 160 of them. Of these, 88 fall in the first redshift bin, 45 in the
second, 21 in the third, and 6 in the fourth. If the sample of SDP galaxies
with $z_{\rm spec}<0.5$ is representative and about 20\% of the 5021 galaxies
in our sample prior to the removal of $R>0.8$ objects have unrecognized
genuine SDSS counterparts, the low-$z$ contaminants would amount to 14\% of
sources in the first redshift bin, 7.4\% in the second, 3.8\% in the third,
1.8\% in the fourth. On the other hand, it is estimated (Smith et al. 2011a)
that $\approx 5.8\%$ of the 2418 $R>0.8$ identifications are spurious. If
these spurious identifications have the same redshift distribution as the
3469 sources not identified with $R>0.8$, we would be missing about 4\% of
sources in each redshift bin. Finally, we have to take into account the
incompleteness of the SDP catalog. This can be done using the flux density
dependent correction factors given in Table 2 of Rigby et al. (2011).

The estimated LFs, taking into account all the corrections described above,
are presented in Figs.~\ref{fig:LF100}--\ref{fig:LF100evo} and in
Tables~\ref{tab:lf100} and \ref{tab:lf250}. The upper scale in these figures
displays the SFR corresponding to the $100$ or $250\,\mu$m luminosity for the
SMM J2135-0102 calibration giving
\begin{equation}
{L_{100\mu\rm m}\over {\rm W~Hz}^{-1}}=5.9\times 10^{23}\,{\mathrm{SFR}\over
M_{\odot}~{\rm yr}^{-1}},
\end{equation}
while for $L_{250\mu\rm m}$ the coefficient is $1.4\times 10^{23}$. Since for
galaxies with intense star formation the rest-frame dust emission peaks in
the range $\lambda\approx 90-100\, \mu$m, the $100\, \mu$m luminosity is a
good estimator of the SFR.

Note that the flattening of the LFs at the lowest luminosities may be, at
least in part, due to the overestimate of the accessible volume yielded by
the $1/V_{\rm max}$ estimator for objects near to the flux limit, pointed out
by Page \& Carrera (2000; see also Eales et al. 2009). On the other hand, the
Page \& Carrera (2000) estimator holds under the assumption that the
luminosity function varies little within the luminosity bin, while in our
case it is very steep over most of the luminosity range.

Fig.~\ref{fig:LF100sed} illustrates the effect of varying the template SED,
after correcting for the median offsets in the redshift estimates highlighted
by Fig.~\ref{fig:zspec}. The differences are quite limited. Such a
\emph{stability} of the LF estimates follows from some favorable
circumstances. First, the large numbers of sources in each redshift bin
smooth out the effect of errors on redshift estimates. Second, for the
redshift range considered here ($1 \la z\la 4$) one of the SPIRE wavelengths
is always sampling directly the rest-frame SED in the range $100\la
\lambda\la 125\, \mu$m, implying that $K$-corrections (and the related
uncertainties) are minimal. Third, the LFs are only moderately sensitive to
the uncertainty on redshift estimates. This can be shown as follows. The
luminosity at a rest-frame frequency $\nu$ is related to the flux density at
the observed frequency $\nu_{\rm obs}$ by
\begin{equation}
L_{\nu}=f_{\nu_{\rm obs}}\,{4\pi\, D^2_L(z)\over (1+z)\times l}~,
\end{equation}
where
\begin{equation}
l={L_{\nu_{\rm obs}\, (1+z)}\over L_{\nu}}.
\end{equation}
Since, in our case, $\nu\approx \nu_{\rm obs}\,(1+z)$, $l\approx 1$.
Then it is easily checked that, in the redshift range of interest here,
$L_{\nu} \propto (1+z)^3$, so that
\begin{equation}
\Delta \log L\approx {3\over \ln(10)}\,\left({\Delta z\over 1+z}\right).
\end{equation}
The LF, $\phi$, is computed by weighting each galaxy by the inverse of the
maximum accessible volume (Schmidt 1968), i.e., $\phi \propto 1/V$. At $z\ga
1$ we have, roughly, $V\propto (1+z)^{2.8}$, whence
\begin{equation}
\Delta \log \phi \approx -1.2\, \left({\Delta z\over 1+z}\right).
\end{equation}
Since $\Delta z/(1+z)\approx 0.2$ the error on the LF determination due to
uncertainty in the redshift estimate amounts to around $0.25$ dex. The
overall uncertainty is the sum in quadrature of the statistical error and of
the error coming from the uncertainty on the redshift estimate. As shown in
Fig.~\ref{fig:LF100evo}, the correction for the Eddington bias is rather
small compared to the overall uncertainty of the LF.

In Figs.~\ref{fig:LF100} and \ref{fig:LF250} our LF estimates are compared
with those by Gruppioni et al. (2010) and Eales et al. (2010b), respectively.
The latter authors adopted a grey body SED with $T=26$ K and a dust
emissivity index $\beta =1.5$. The assumed dust temperature is possibly more
appropriate for relatively low SFR/far-IR luminosity galaxies while typical
dust temperatures of SPIRE-detected $z\approx 2$ galaxies are somewhat higher
(Chapman et al. 2010; Amblard et al. 2010). The different choice for the SED
has a substantial impact on the $K-$correction and therefore on the estimated
rest-frame luminosity. For example, for a galaxy at $z\approx 1.5$ and a
given observed $250\,\mu$m flux density, the Eales et al. SED yields a
rest-frame $250\,\mu$m luminosity a factor $\approx 1.5$ higher than that
obtained using either the SMM J2135-0102 or the Arp220 SED. Once we replace
their $K-$correction with ours, the LF estimates by Eales et al. (2010b) are
found to be in good agreement with ours (see Fig.~\ref{fig:LF250}). The two
estimates are to some extent complementary: the deeper samples used by Eales
et al. have allowed them to reach lower luminosities, while the larger
\textsl{H-ATLAS} SDP area has allowed us to reach higher luminosities at high
redshifts.

Gruppioni et al. (2010) exploited the deep PACS data at $100$ and $160\,\mu$m
in the GOODS-N field ($\sim 150\,\mathrm{arcmin}^2$), obtained as part of the
PACS Evolutionary Probe (PEP) survey, to estimate the $60$ and $90\,\mu$m
rest-frame LFs up to $z\sim 3$. They have used all the available data to
derive the SEDs of their sources.

About $31\%$ of Eales et al. sources have spectroscopic redshifts and almost
all the others have photometric redshifts typically based on nine optical and
near-IR bands. Gruppioni et al. (2010) have spectroscopic redshifts for $\sim
70\%$ of their sources; for the remaining $\sim 30\%$ they have obtained
photometric redshifts from multifrequency data. The agreement between our LFs
and theirs is an additional confirmation that photometric redshifts derived
from sub-mm photometry are good enough for the present purpose.

In the sampled luminosity range, the LFs exhibit an exponential fall off and
a substantial luminosity evolution at least up to $z\approx 2.5$, while a
weaker evolution at higher $z$ is indicated (see Fig.~\ref{fig:LF100evo}).
The $250\,\mu$m luminosity corresponding to $\log(\phi/{\rm Mpc}^{-3})=-5$ is
$\log(L_{250}/{\rm W\,Hz}^{-1}) = 26.02$, $26.16$, $26.32$, and $26.38$ for
$z\approx 1.4$, $1.8$, $2.2$, $3.2$, respectively. At $100\,\mu$m the
corresponding luminosities are $\log(L_{100}/{\rm W\,Hz}^{-1}) = 26.65$,
$26.78$, $26.94$, and $27.01$. These results are consistent with those based
on PEP survey data (Gruppioni et al. 2010), which have poorer statistics at
high $z$. We remark that our data supplemented with those by Gruppioni et al.
(2011) and Eales et al. (2010b) allow the determination of the LFs over only
one order of magnitude in luminosity for $1\la z\la 2$, and over only a
factor of about $3$ for $z\ga 2$.

\section{Clues on star-formation timescales in massive halos}\label{sect:disc}

We now discuss how the LFs of high redshift star forming galaxies at sub-mm
wavelengths and the counts in sub-mm and mm bands concur with their
clustering properties in probing the process of star formation in the
progenitors of massive ETGs.

\subsection{Clustering properties and host halo masses of sub-mm galaxies}\label{sect:vdisp}

Several lines of evidence indicate that high-$z$ (sub-)mm bright galaxies are
strongly clustered (Blain et al. 2004; Farrah et al. 2006; Magliocchetti et
al. 2007; Viero et al. 2009; Maddox et al. 2010; Hall et al. 2010; Cooray et
al. 2010; Dunkley et al. 2010; Amblard et al. 2011; Planck Collaboration
2011). The large-scale clustering power-spectrum probes the relation between
the distribution of visible sources and that of dark matter halos, which is a
very sensitive function of the halo mass scale (e.g. Matarrese et al. 1997).

The study of the angular correlation function of \textsl{H-ATLAS} SDP
galaxies (Maddox et al. 2010) did not detect significant clustering for the
$250\,\mu$m selected sample, while the clustering signal was found to be
strong (although with large uncertainties) for samples selected at 350 and
$500\,\mu$m. The measurements are consistent with the idea that sub-mm
sources consist of a low redshift population of moderately star-forming
galaxies and a high redshift population of highly clustered star-bursting
galaxies. The former, known to be weakly clustered (e.g. Madgwick et al.
2003; Guo et al. 2011; van Kampen et al. 2011), dominate the $250\,\mu$m
selected sample, while the selection at longer wavelengths emphasizes
strongly clustered galaxies with intense star-formation activity at $z\approx
2-3$, with typical halo masses $\approx 10^{13}\, M_{\odot}$, which are
interpreted as the ancestors of present day massive ellipticals (Negrello et
al. 2007).

\subsection{Timescales of sub-mm and UV bright phases}

The bright end of the LFs provides information on the average duty cycle of
the star formation in the large halos indicated by the clustering analysis.
This can be illustrated using a simple model (hereafter referred to as the
`toy model') relying on the following assumptions:
\begin{itemize}

\item Most of the star formation occurs soon after the fast collapse
    phase of the halo, as identified by Zhao et al. (2003) and by several
    subsequent works (e.g., Diemand et al. 2007, Genel et al. 2010, Wang
    et al. 2011).

\item The halo formation rate is given by the positive term of the
    derivative with respect to the cosmic time  of the Sheth \&
    Tormen (1999, 2002) cosmological mass function [eq. (1) of Lapi et
    al. (2006)]. This was shown to be a good approximation at the
    redshifts ($z\ga 1$) and for the halo masses ($M_H\ga 3\times
    10^{12}\, M_{\odot}$) of interest here (e.g., Haehnelt \& Rees 1993;
    Sasaki 1994; see also the discussion by Lapi et al. 2006).

\item During the main episode of star formation, the SFR is proportional
    to the halo mass, i.e., $\mathrm{SFR}\propto M_{H}$, with a small
    scatter.

\item The duration $\Delta t_{\rm sf}$ of the main star formation
    episode, before quenching by the AGN feedback, is roughly constant
    for the considered range of halo masses ($3\times 10^{12}\,
    M_{\odot}\la M_H\la 3\times 10^{13}\, M_{\odot}$) and shorter than
    the expansion timescale at the source redshift.

\end{itemize}
This very simple model provides a good fit to the data (Figs.~\ref{fig:LF100sed} and
\ref{fig:LF250})
if $\Delta t_{\rm sf}\approx 7\times 10^{8}$ yr and:
\begin{equation}\label{eq:SFR}
\mathrm{SFR}=35\left(\!{M_H\over 10^{12}\, M_{\odot}}\!\right)\left({1+z\over
2.5}\right)^{2.1}\, M_{\odot}\,{\rm yr}^{-1},
\end{equation}
with a rather small ($\pm 35\%$) scatter. We have used Eq.~(\ref{eq:conv})
with the calibration appropriate for SMM J2135-0102 to pass from star
formation rate to FIR luminosity, and the SED of this object to go from the
FIR luminosity to monochromatic luminosities. The fit would significantly
worsen if the halo mass range \emph{associated to individual galaxies}
extends above $M_H\ga 3\times 10^{13}\, M_{\odot}$. This suggests that the
star formation becomes very inefficient in the most massive halos, possibly
due to the cooling time becoming longer than the expansion time. The bright
end of the LFs is quite sensitive to the value of $\sigma_8$; e.g., raising
$\sigma_8$ from our fiducial value of $0.81$ to $0.9$ would increase by a
factor $\approx 2$ the $250\, \mu$m LF at $z\approx 2.5$ and $L_{250\mu\rm
m}\approx 3\times 10^{26}$ W Hz$^{-1}$. Interestingly, the duration of the
main star-formation episode yielded by the toy model matches that inferred
from the observed $\alpha$-enhancement of massive local ETGs (see
\S\,\ref{sect:intro}).

A rough estimate of the mass in stars at the end of the main star-formation
episode, for given halo mass and redshift, can be obtained multiplying the
appropriate SFR by its duration $\Delta t_{\rm sf}\approx 7\times
10^{8}\,$yr, and correcting for the fraction of stellar mass returned to the
interstellar medium. For a Chabrier IMF the latter amounts to $\approx 30\%$
after $0.1$ Gyr from a burst of star formation, and increases to $\approx
40\%$ after $1$ Gyr and to $50\%$ after several Gyrs. Combining with the halo
mass function at that redshift we get an estimate of the stellar mass
function of ETGs. A comparison with observational determinations at different
redshifts is shown in Fig.~\ref{fig:MF}. The agreement is remarkably good,
considering the crudeness of the approach. This suggests that the toy model
captures the basic aspects of the star formation in massive galaxy halos at
high redshift.

A comparison of our LFs with the UV ones at $z\approx 2-3$ (Sawicki \&
Thompson 2005; Reddy \& Steidel 2009) for galaxies with comparable SFRs shows
that the UV space densities are a factor $\ga 100$ lower, implying that the
UV bright, dust-free phase is much shorter than the far-IR bright phase. The
far steeper slope of the UV LF, compared with the sub-mm ones, for
$\mathrm{SFR}\sim$ hundreds $M_{\odot} \mathrm{yr}^{-1}$ implies that, for
the corresponding range of galaxy masses, the UV-bright phase is shorter for
larger SFRs, i.e. for more massive galaxies, as predicted by Mao et al.
(2007).

If, as implied by most current semi-analytical models (Granato et al. 2001,
2004; Croton et al. 2006; Hopkins et al. 2008; Somerville et al. 2008), the
black hole growth is linked to star-formation, most of the present-day black
hole mass should have been accreted by the end of the star-forming phase (see
Bonfield et al. 2011). Therefore, the local ratio
$M_{\bullet}/M_{\star}\approx 2\times 10^{-3}$ (Marconi \& Hunt 2003) between
the black hole mass $M_{\bullet}$ and the stellar mass $M_{\star}$ of the
host ETG should be already in place then (but see also McLure et al. 2006 and
Peng et al. 2006). If so, the stellar mass function translates immediately
into a black hole mass function which, in turn, can be translated into a
bolometric luminosity function assuming that the black holes, immediately
before being switched off, emit at the Eddington limit for a time $t_{\rm
vis}$. We get:
\begin{equation}\label{eq:Lbol}
{L_{\rm BH,bol} \over L_\odot} \approx 79\, {M_\star \over M_\odot}~.
\end{equation}
The corresponding $B-$band luminosity is $L_B=0.1(10/k_{\rm bol})L_{\rm
BH,bol}$, where $k_{\rm bol}$ is the bolometric correction (Marconi et al.
2004; Hopkins et al. 2007). We can then compare the comoving density of
galaxies with that of the associated QSOs. For example, at $z\approx 2$ there
are $\approx 2.7\times 10^{-5}$ sub-mm galaxies per $\mathrm{Mpc}^{3}$ with
$M_\star \ga 2\times 10^{11}\,M_\odot$, while the density of QSOs brighter
than the corresponding luminosity of $L_B \approx 1.6\times 10^{12}\,L_\odot$
($k_{\rm bol}\approx 10$) is $\approx 8\times 10^{-7}\,\mathrm{Mpc}^{-3}$
(Croom et al. 2004). This implies that the optical visibility time of the
QSOs is about a factor of $30$ shorter than the duration of the sub-mm bright
phase, in agreement with independent estimates (Shankar et al. 2004; Marconi
et al. 2004; see also Lapi et al. 2006).

The physical model by Granato et al. (2001, 2004), further elaborated by Lapi
et al. (2006), complies with the indications coming from the previous
analysis. In this model the star formation in massive galaxies occurs on a
timescale of several $10^8$ yr, is very soon obscured by dust and is stopped
by quasar feedback. The star formation is triggered by the rapid cooling of
the gas within a region with an approximate size $\approx 1/3-1/4$ of the
halo virial radius, i.e., $20-30$ kpc, encompassing about $40\%$ of the total
mass (dark matter plus baryons), and is regulated by the energy feedback from
SNe and AGNs (the latter being relevant especially in the most massive
galaxies). As a result, the star formation is a very inefficient process, as
proved by the fact that on the average only about $5-10\%$ of the baryons in
the Universe are converted into stars (Fukugita \& Peebles 2004), with a
possible maximum of $15-20\%$ for halo masses $\sim 10^{12}\,M_\odot$
(Shankar et al. 2006; Moster et al. 2010). It is worth noticing that the
Negrello et al. (2007) predictions for the lensed sub-mm bright galaxies
follow directly from the LF and redshift distribution yielded by the Lapi et
al. (2006) model. We have updated this model by replacing the Arp220 GRASIL
SED used, e.g., by Negrello et al. (2007) with the SED of SMM J2135-0102. The
resulting LFs are shown in Figs.~\ref{fig:LF100} and \ref{fig:LF250}, labeled
as `Full model'. We stress that the LFs from the Full model are not fits to
the data, but have been computed with the same parameters used by Lapi et al.
(2006; see their Table~1). This model has proven to be successful in
reproducing a wealth of observables, including high redshift quasar LFs.

Note that, in the present framework, sources dominating the high-$z$ LF are
proto-spheroidal galaxies that are in passive evolution at $z\la 1.5-1$ and
therefore do not contribute to the low-$z$ FIR/sub-mm LFs computed, e.g., by
Dye et al. (2010), Vaccari et al. (2010), and Dunne et al. (2011).

An alternative model whereby intense star-formation at $z\approx 2-3$ is not
supported by mergers was proposed by Dekel et al. (2009). In this model,
star-formation is driven by steady cold streams supplying gas at an
approximate rate of
\begin{equation}\label{eq:Dekel}
\dot{M}\approx 52 \left({M_H\over 10^{12}\,
M_{\odot}}\right)^{1.15}\,\left({1+z\over
2.5}\right)^{2.25}~f_{0.165}\,M_{\odot}~{\rm yr}^{-1},
\end{equation}
where $f_{0.165}$ is the baryonic fraction in the halos in units of the
cosmological value, $f_b=0.165$. The dependence of the cold gas inflow rate,
$\dot M$, on $M_H$ and $z$ are similar to those implied by our toy model
[Eq.~(\ref{eq:SFR})], implying that the SFR must be roughly proportional to
$\dot M$. To account for the SFR indicated by the \textsl{Herschel} data, the
star-formation efficiency must be very high, $\sim 70\%$ for $M_H\approx
10^{12}\, M_{\odot}$ and $z\approx 1.5$. For comparison, the mean cosmic
star-formation efficiency is $\Omega_{\rm star}/\Omega_{\rm baryon}\approx
(5.8 \pm 1.1)\%$ (Fukugita \& Peebles 2004; Komatsu et al. 2011). In halos
more massive than $\approx 10^{13}\,M_\odot$ cold streams are suppressed by
shock heating.

\section{(Sub)-mm counts and redshift distributions}\label{sect:counts}

Figs.~\ref{fig:counts250}, \ref{fig:counts350}, and \ref{fig:counts500}
compare the predictions of the 'full model' with the observed counts in
\textsl{Herschel}/SPIRE bands. In these figures the contributions to the
counts of proto-spheroidal galaxies have been complemented with those of
normal late-type and starburst galaxies computed by Negrello et al. (2007).
As for massive proto-spheroidal galaxies, the main difference with the counts
in the latter paper comes from having replaced the Arp220 SED yielded by
GRASIL with that of SMM J2135-0102. The contribution from lensed
proto-spheroidal galaxies has been estimated using the amplification
distribution of Perrotta et al. (2003) and Negrello et al. (2007).

An interesting prediction of the model is that massive proto-spheroidal
galaxies dominate the (sub-)mm counts over a limited flux density range
(about a decade). At $250\,\mu$m the Euclidean normalized differential counts
of these objects peak at $\approx 30$ mJy, i.e. roughly at the detection
limit of the \textsl{H-ATLAS} survey, and sink down rapidly at fainter fluxes
where the contribution of starburst galaxies takes over and accounts for the
results of the $P(D)$ analysis by Glenn et al. (2010). Since the
proto-spheroidal galaxies are mostly at $z\ga 1.5$ while late-type/starburst
galaxies are mostly at $z\la 1.5$, the model implies that the redshift
distribution drifts towards lower redshifts as we go fainter. This is
not in contradiction with the finding by Bourne et al. (2011) that the
contribution of $z<0.28$ galaxies to the cosmic infrared background is $<5\%$
in the SPIRE bands.

The model accurately fits the source counts from $250\,\mu$m to $850\,\mu$m
(Figs.~\ref{fig:counts250}$-$\ref{fig:counts850}). The observed counts at 1.1
mm (Fig.~\ref{fig:counts1100}) span the peak of the Euclidean normalized
counts, while those at 1.4 and 2 mm (Figs.~\ref{fig:counts1400} and
\ref{fig:counts2000}) only cover the brightest tail of the counts, dominated
by strongly lensed sources (apart from radio sources). The model somewhat
overestimates the counts of strongly lensed galaxies at 1.4 and 2 mm, and the
most recent counts at 1.1 mm. This may suggest that higher-$z$ galaxies, that
yield larger and larger contributions to the bright counts at increasing mm
wavelengths, have SEDs slightly colder than SMM J2135-0102 and closer to that
of G15.141.

The flux density range over which massive proto-spheroidal galaxies dominate
the counts increases somewhat with increasing wavelength
(Figs.~\ref{fig:counts250}$-$\ref{fig:counts2000}), reflecting the increase
of the redshift range through which the strongly negative $K-$correction
makes the flux corresponding to a given luminosity essentially independent of
distance. The broadening of the peak in the Euclidean normalized counts of
proto-spheroidal galaxies is not very large, however, because massive halos
become increasingly rare at high $z$. At all wavelengths the bright counts of
proto-spheroidal galaxies drop down very steeply, reflecting the exponential
decline of the halo mass function. This rapid fall-off, borne out by the
data, can hardly be accounted for by phenomenological models that evolve the
LF of local populations of dusty galaxies backwards in time: spheroidal
galaxies are essentially in passive evolution since $z\approx 1.5$ and are
therefore not represented in local LFs at far-IR to mm wavelengths.

The physical model of Lacey et al. (2010) works pretty well at $850\,\mu$m
but does not correctly predict the rapid fall-off of the counts above
$\approx 30$ mJy at $250\,\mu$m, $350\,\mu$m and $500\,\mu$m
(Figs.~\ref{fig:counts250}$-$\ref{fig:counts850}). The decline of the counts
is most easily understood if the duration of the most active star-formation
phase in massive halos is relatively short, as also required by the observed
$\alpha$-enhancement (see \S\,\ref{sect:intro}). However, in merger-driven
evolutionary models, such as the one by Lacey et al. (2010), the star
formation does not truncate after $\la 1$ Gyr. At $850\,\mu$m consistency
with the data can be recovered if sub-mm bright galaxies contain large
amounts of cold dust, but this leads to $250\,\mu$m, $350\,\mu$m and
$500\,\mu$m counts less steep than observed.

The fact that using, for high-$z$ galaxies, the SMM J2135-0102 SED we are
able to simultaneously reproduce the source counts over a broad wavelength
range indicates that this SED is indeed typical for sub-mm bright galaxies.
In fact the results on the counts vary substantially if we use different SEDs
(see Fig.~\ref{fig:counts_comp}). Almost by construction, in all cases the
$250\,\mu$m (the main selection wavelength) counts are accurately reproduced.
But if, for example, we use the SED of G15.141 instead of the SMM J2135-0102
one, the longer wavelength counts are somewhat underestimated. The situation
is better, but still not as good as for our preferred SED, if we use the
Arp220 SED. The quite cold ($T_d=26$ K) SED used by Eales et al. (2010b)
leads to counts substantially in excess of the observed ones. Thus the
`colors' of the counts  constrain the SEDs of high redshift star forming and
in particular on the abundance of galaxies with cold dust temperatures. As
pointed out by Hwang et al. (2010) apparently very cold values of $T_d$ could
be caused by an overestimation of the sub-millimeter fluxes due to blending
problems or inappropriate single-temperature SED fitting.

Fig.~\ref{fig:zdist_comp} compares the redshift distributions of $z>1$
\textsl{H-ATLAS} SDP sources detected at $S>35$, $40$, $48$ mJy at $250$,
$350$ and $500\,\mu$m, respectively, with those predicted by the model. The
effective flux correction factors tabulated by Clements et al. (2010) have
been applied for this comparison. Since the redshift distributions are a key
ingredient for the successful prediction of the abundance of strongly lensed
galaxies and of their redshift range (Negrello et al. 2010), they cannot be
badly wrong. Thus, the agreement between our estimates and model predictions
is a confirmation of the global consistency of our results.

The shift in the redshift distribution of sub-mm bright galaxies to higher
redshifts with increasing selection wavelength happens because the strongly
negative $K-$correction extends to larger and larger redshifts and the
lensing probability increases with $z$. For example, we expect that, after
removing local objects that are found to be only $\approx 1/3$ of the total
(Vieira et al. 2010), the redshift distribution of sources with $S_{\rm
1.4}\ga 10$ mJy has a broad peak around $z\approx 4.5$. Using the SMM
J2135-0102 SED we also find that the observed $1.4-2$ mm spectral indices are
correlated with $z$: for $z\la 4.5$, $\alpha_{\rm 1.4}^{2}\ga 3.2$, while for
$z\ga 4.5$, $\alpha_{\rm 1.4}^{2}\la 2.6$. Thus the value of this spectral
index may be a rough redshift indicator.

In Fig.~\ref{fig:background} we illustrate the contribution of
proto-spheroids to the extragalactic infrared background intensity (Lagache
et al. 1999; Hauser \& Dwek 2001) according to our physical model.  At
$\lambda\ga 850\,\mu$m the background is almost entirely accounted for by
high-redshift proto-spheroids, while these contribute about $65\%$ and about
$50\%$ of the background at $\lambda\approx 500\, \mu$m and $\lambda\approx
250\, \mu$m, respectively.

\section{Conclusions}\label{sect:concl}

We have exploited the \textsl{H-ATLAS} SDP survey data to investigate the
evolution of the 100 and $250\,\mu$m LFs of bright star-forming galaxies
($\mathrm{SFR} \ga 10^2\, M_{\odot}\,\mathrm{yr}^{-1}$) at $z\ga 1$.
Redshifts have been estimated using 3 SED templates representative of the
range of well measured SEDs of galaxies with such a high star formation. The
rms uncertainties on redshift estimates have been assessed comparing our
redshift estimates with spectroscopic redshift measurements for 39
\textsl{H-ATLAS} galaxies at $z>0.5$ as well as by means of simulations. Both
methods yield rms values of $\Delta z/(1+z)\approx 0.2$ or smaller. The LFs
derived using the redshift estimates based on each of the 3 SED templates are
very similar to each other. The uncertainties due to the spread of redshift
estimates are added in quadrature to Poisson errors to compute the global
uncertainties on the LFs. Our LF estimates are in close agreement (in the
common redshift and luminosity range, after applying the same
$K-$corrections) with those at $250\,\mu$m by Eales et al. (2010b), based on
a substantial fraction of spectroscopic redshifts complemented with
photometric redshifts of optical/near-IR counterparts, as well with those at
$90\,\mu$m by Gruppioni et al. (2010), based on an even higher fraction of
spectroscopic redshifts. This is a further confirmation that our redshift
estimates are sufficiently accurate to allow reliable estimates of the LFs.

The SED of SMM J2135-0102 was found to perform significantly better than the
others in the tests we made, at least up to $z\approx 3$, and was therefore
adopted as our reference. Remarkably, this SED allowed us to simultaneously
fit the counts over a broad wavelength range.

We find (see Fig.~\ref{fig:LF100evo}) a significant luminosity evolution at
least up to $z\approx 2.5$ while the luminosity function shows a
modest variation between the last two redshift bins, centered at $z\approx
2.2$ and $z\approx 3.2$, respectively, even though the corresponding time
interval is $\approx 1\,$Gyr, substantially larger than the time interval
($\approx 0.65\,$Gyr) between the central redshifts of the second and third
bin ($z\approx 1.8$ and $\approx 2.2$). This is consistent with the results
based on PEP survey data (Gruppioni et al. 2010).  We show that the evolution
of the LF reflects that of the halo formation rate if, for the very massive
galaxies represented in our sample, the star-formation rate obeys a simple
relationship with halo mass (estimated from the clustering properties) and
redshift [Eq.~(\ref{eq:SFR})] and the lifetime of the main star-formation
phase is $\Delta t_{\rm sf}\approx 7\times 10^{8}$ yr, consistent with the
constraint coming from the $\alpha$-enhancement observed in the most massive
ETGs.

The stellar mass function resulting from the derived star-formation rates and
timescale $\Delta t_{\rm sf}$ nicely fits the observed stellar mass function
of passively evolving ETGs at $z\approx 1-2$.

A comparison of sub-mm and UV LFs indicates that the UV visibility time of
massive galaxies is much ($\ga 100$ times) shorter than the duration of the
sub-mm bright phase ($\Delta t_{\rm sf}\approx 7\times 10^8$ yr). This
implies that dust  forms very rapidly after the onset of the main episode of
star formation, either in the ISM (Draine et al. 2009; Dunne et al.
2011) or by effect of Type-II SNae (Dunne et al. 2003, 2009; Matsuura et al.
2011).

As discussed by Mao et al. (2007), a longer UV bright phase
is expected for less massive galaxies. This prediction is supported by the
flatter slope of the far-IR LF, compared to the UV one, for
$\mathrm{SFR}\sim$ hundreds $M_{\odot} \mathrm{yr}^{-1}$, implying a decrease
with increasing galaxy mass of the ratio of UV to far-IR space densities,
i.e. of the ratio between UV and far-IR lifetimes.

In the same vein we find that the duration of the optically bright QSO phase
is of order $1/30$ of $\Delta t_{\rm sf}$, i.e., of about $2-3\times 10^7$
yr, consistent with independent evidences. On the other hand, the bright end
of the QSO LF is somewhat less steep than that of sub-mm galaxies. This is
not surprising because, although the evolution of the SFR and of accretion
into the central black hole may well be linked, the relationship is mediated
by several steps that introduce a substantial scatter that, coupled with the
curvature of the LF, flattens its bright tail.

The $100$ and $250\,\mu$m LFs at different redshifts are quite well
reproduced by the physical model of ETG formation and evolution by Granato et
al. (2001, 2004), further elaborated by Lapi et al. (2006), \emph{without any
adjustment of the parameters}. As discussed in these papers, the
model is built in the framework of the standard hierarchical clustering
scenario. Many simulations (e.g., Zhao et al. 2003, Diemand et al. 2007,
Genel et al. 2010, Wang et al. 2011) have shown that the growth of a halo
occurs in two different phases: a first regime of fast accretion in which the
potential well is built up by the sudden mergers of many clumps with
comparable masses; and a second regime of slow accretion in which mass is
added in the outskirts of the halo, only occasionally affecting the central
region where the galactic structure resides. According to the model, the fast
accretion phase triggers a burst of star formation that, in massive halos at
$z\ga 1$, starts an evolutionary sequence that can be summarized as follows.
There is an early, short phase of (almost) dust-free star formation when the
galaxy shines as a bright UV source. It is followed by a dust-enshrouded star
formation phase when the galaxy shines in the far-IR/sub-mm range. The
duration of both the UV bright phase and the far-IR/sub-mm bright phase is
shorter for the most massive galaxies, with the highest SFRs; for these
objects the UV phase lasts $\la 10^7$ yr and the far-IR/sub-mm phase lasts
$<10^9\,$yr. Then there is a phase, lasting several $10^7$ yr, when the
nucleus shines as a bright QSO after having swept away most of the
interstellar gas and dust. Finally, passive evolution of the stellar
populations follows, and the galaxy evolves into a local ETG.

According to this model, star-forming proto-spheroidal galaxies
account for a substantial fraction of the cosmic infrared background (see
Fig.~\ref{fig:background}) and dominate the cosmic SFR at $z>1.5$, while at
lower $z$ the SFR is dominated by late-type (normal and starburst) galaxies.
This model was the basis for the successful predictions the sub-mm counts of
strongly lensed galaxies by Perrotta et al. (2003) and Negrello et al.
(2007). It also accurately reproduced the epoch-dependent galaxy
luminosity functions in different spectral bands, as well as a variety of
relationships among photometric, dynamical and chemical properties, as shown
in previous papers (see Table 2 of Lapi et al. 2006 and additional results,
especially on the galaxy chemical evolution, in Mao et al. 2007 and Lapi et
al. 2008).

\begin{acknowledgements}
The \textsl{Herschel}-ATLAS is a project with \textsl{Herschel}, which is an
ESA space observatory with science instruments provided by European-led
Principal Investigator consortia and with important participation from NASA.
The \textsl{H-ATLAS} website is http://www.h-atlas.org/

The work has been supported in part by ASI/INAF agreement n. I/009/10/0 and
by INAF through the PRIN 2009 ``New light on the early Universe with sub-mm
spectroscopy''. We thank the referee for helpful comments and
suggestions, and Cedric Lacey who provided in tabular form the sub-mm counts
yielded by the Lacey et al. (2010) model. A. Lapi acknowledges useful
discussions with A. Cavaliere, G.L. Granato, P. Salucci, L. Silva and F.
Shankar, and thanks SISSA and INAF-OATS for warm hospitality.
\end{acknowledgements}

\clearpage
\begin{figure}
\plotone{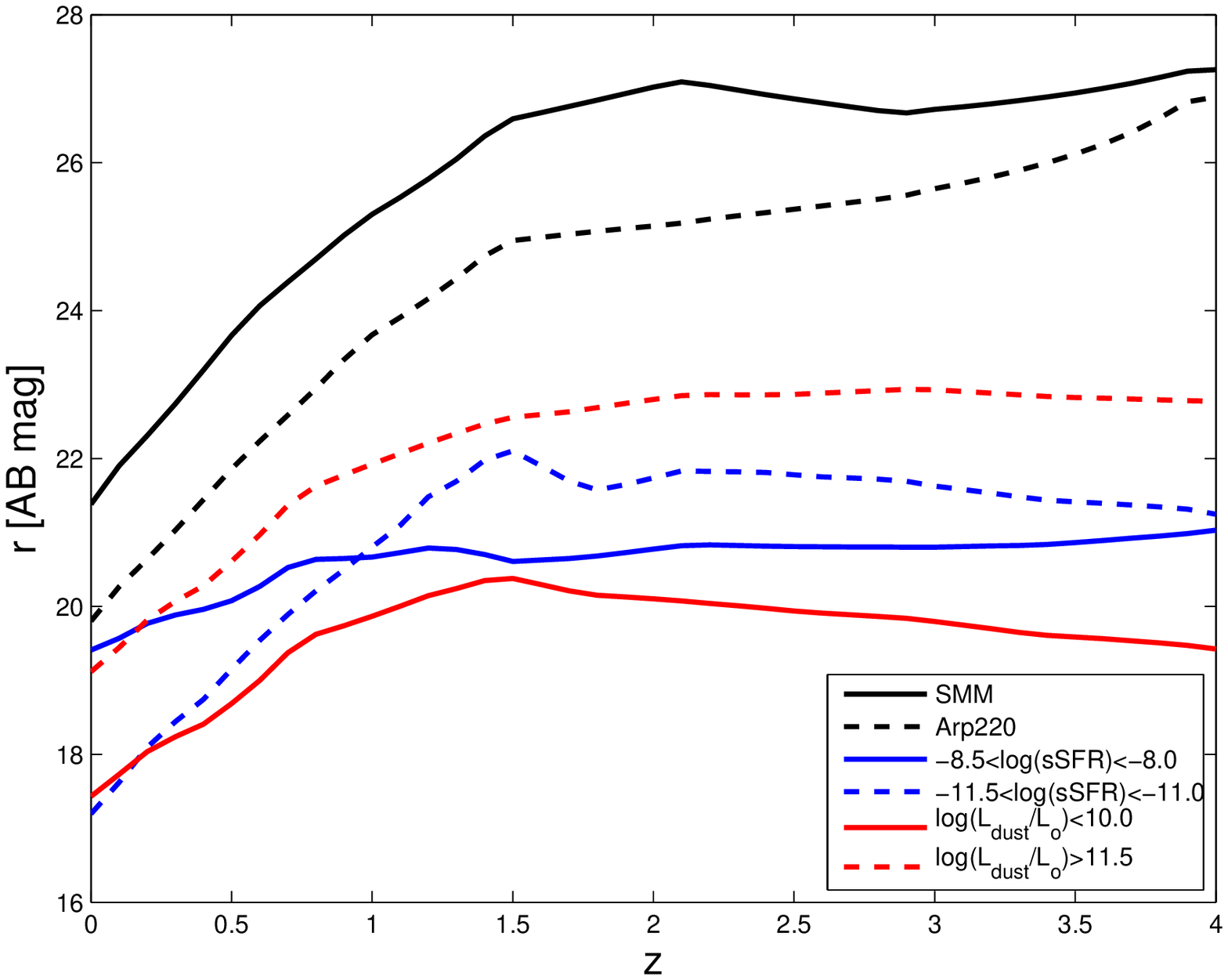}
\caption{Optical (SDSS $r-$band) magnitudes as a function of redshift
for several SEDs, normalized to $S_{250\mu\rm m}=33.5\,$mJy, the $5\sigma$
detection limit for \textsl{H-ATLAS} SDP galaxies. The red curves refer to
the mean SEDs of optically identified $z<0.5$ SDP galaxies in the lowest
[$9.5 < \log(L_{\rm dust}/L_\odot) < 10$; solid line] and in the highest
[$11.5 < \log(L_{\rm dust}/L_\odot) < 12$; dashed line] luminosity bins of
Smith et al. (2011b).  The blue curves refer to the same sample by Smith et
al. (2011b) but for the lowest [$-11.5 < \log({\rm sSFR})< -11$; dashed] and
for the highest [$-8.5 < \log({\rm sSFR}) < -8$; solid] specific star
formation rate (in yr$^{-1}$) bins. The dashed black line refers to the SED
of Arp220 (a local ULIRG) and the black solid line to that of SMM J2135-0102
(`The Cosmic Eyelash' at $z\approx 2.3$). As explained in \S~2, the figure is
meant to illustrate that the high-$z$ \textsl{H-ATLAS} sources must have
global SEDs different from the low-$z$ ones. The latter are relatively cool
and unobscured, while the high-$z$ population is dominated by more obscured
objects with higher star-formation rates and warmer dust
temperatures.}\label{fig:rmag}
\end{figure}

\clearpage
\begin{figure}
\plotone{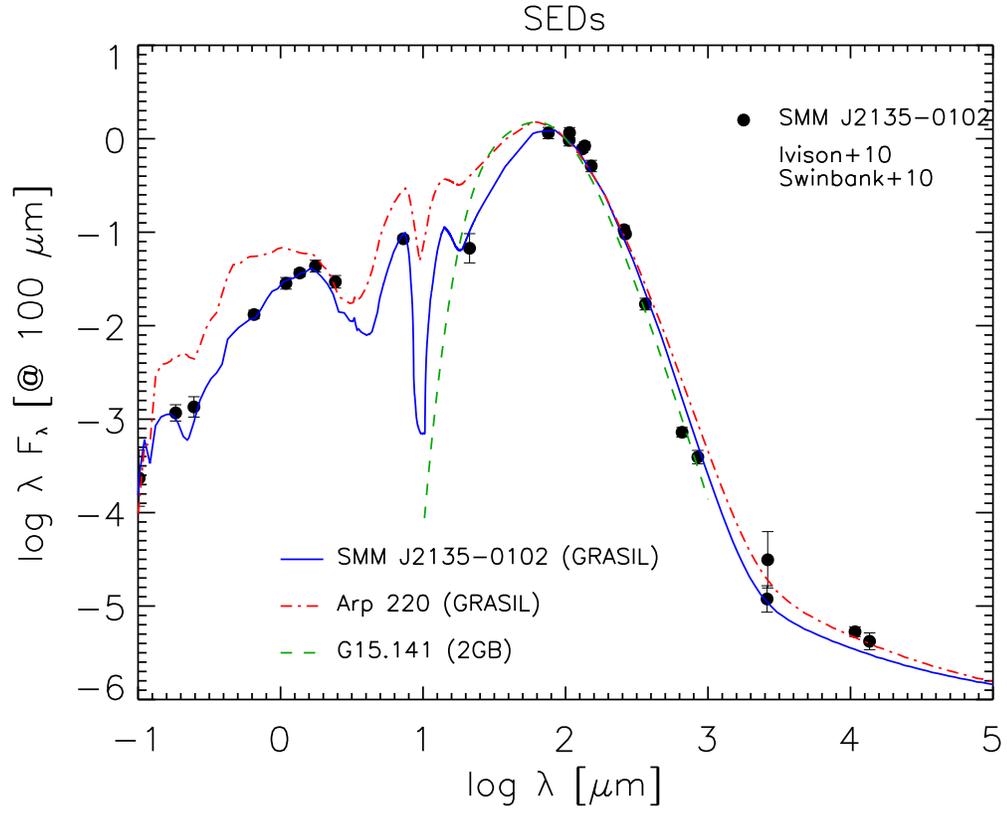}
\caption{The SED of SMM J2135-0102 (`The Cosmic Eyelash') as modeled by
GRASIL (blue solid line); data are from Ivison et al. (2010) and Swinbank et
al. (2010). The SED of Arp220 (red dot-dashed line) as modeled by GRASIL, and
that of G15.141 modeled as a double-temperature grey-body (green dashed line;
see text for details) are also shown for comparison. All SEDs are in the
rest-frame and are normalized at $100\,\mu$m.}\label{fig:SED}
\end{figure}

\clearpage
\begin{figure}
\plotone{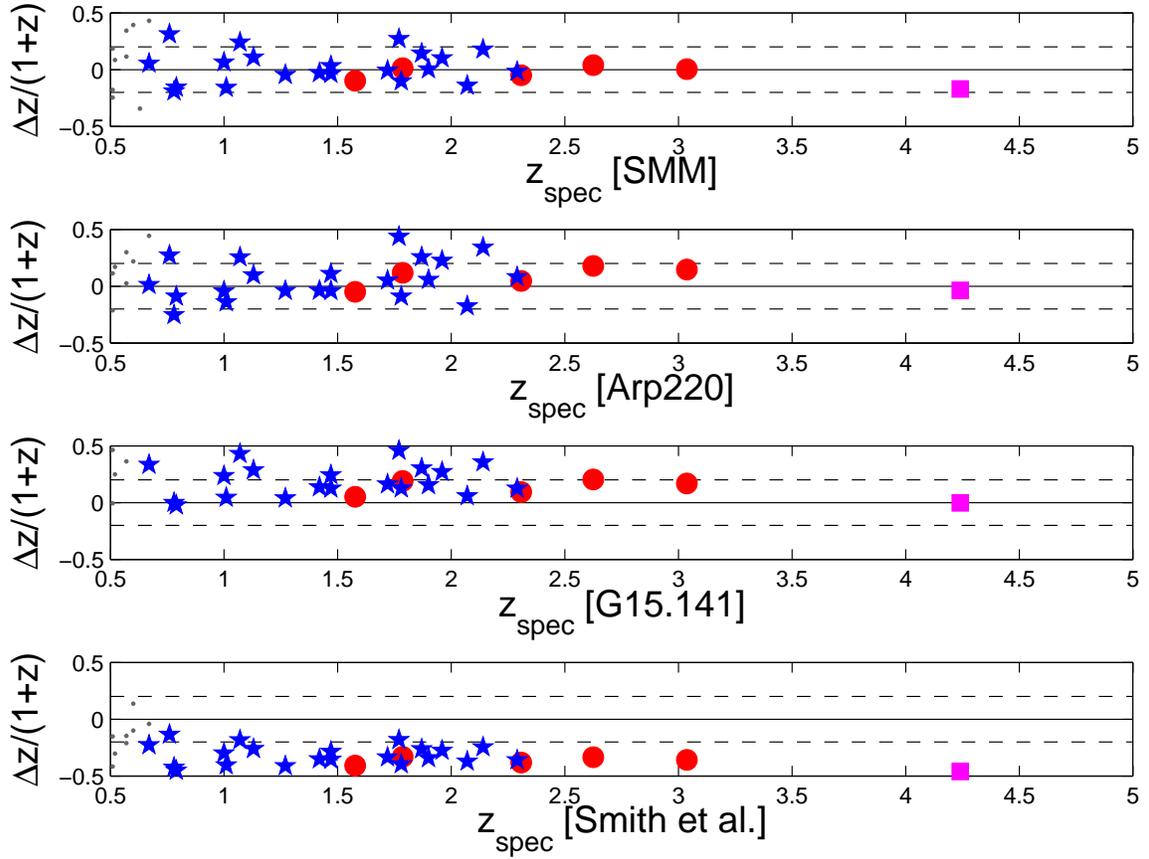}
\caption{Photometric redshift estimates based on the SEDs of (from top to
bottom) SMM J2135-0102, Arp220, G15.141, and the average low-$z$ SED from
Smith et al. (2011) compared in terms of the quantity $\Delta z/(1+z)\equiv
(z_{\rm phot}-z_{\rm spec})/(1+z_{\rm spec})$ with spectroscopic measurements
for the strongly lensed galaxies of Negrello et al. (2010; red filled
circles) and Cox et al. (2011; purple filled square), and for the
\textsl{H-ATLAS} SDP galaxies with $z>0.5$ of Bonfield et al. (2011; blue
asterisks) and Smith et al. (2010; black dots). The dashed lines correspond
to a $20\%$ difference.}\label{fig:zspec}
\end{figure}

\clearpage
\begin{figure}
\plotone{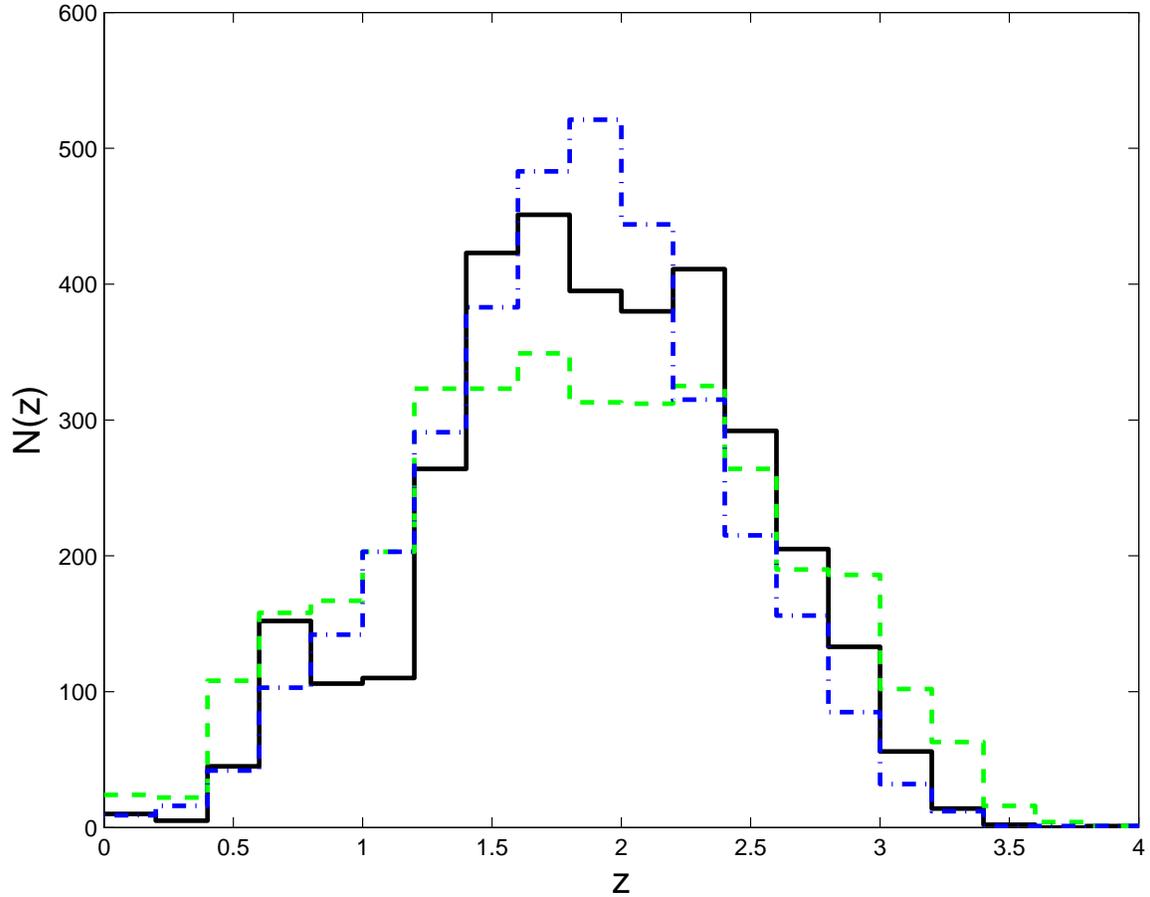}
\caption{Comparison of the redshift distributions of sources in our reference
sample ($S_{250\mu\rm m}>35\,$mJy, $S_{350\mu\rm m}>3\sigma$, and no $R>0.8$
optical identifications; 3469 sources, i.e. $\approx 50\%$ of the 6876
\textsl{H-ATLAS} SDP galaxies; the selection has mostly excluded the
$z<0.5$ sources, and in fact there is another peak in the redshift
distribution of the total SDP sample at $z<0.5$, see Smith et al. 2011a)
estimated using the SEDs of Arp 220 (green dashed line), G15.141 (blue
dot-dashed line), and of SMM J2135-0102 (black solid line), correcting for
the offsets estimated from Fig.~\protect\ref{fig:zspec}. For our preferred
SED, the one of SMM J2135-0102, 3093 are estimated to be at $z>1$. Correcting
for the contamination by cold low-$z$ galaxies with genuine $r<22.4$
counterparts not identified with $R>0.8$ (see \S\,\ref{sect:LF}), we estimate
that the fraction of SDP galaxies with $S_{250\mu\rm m}\ge 35\,$mJy (6100 in
total) that are at $z>1$ is of $\approx 45\%$.} \label{fig:zdist}
\end{figure}

\clearpage
\begin{figure}
\plotone{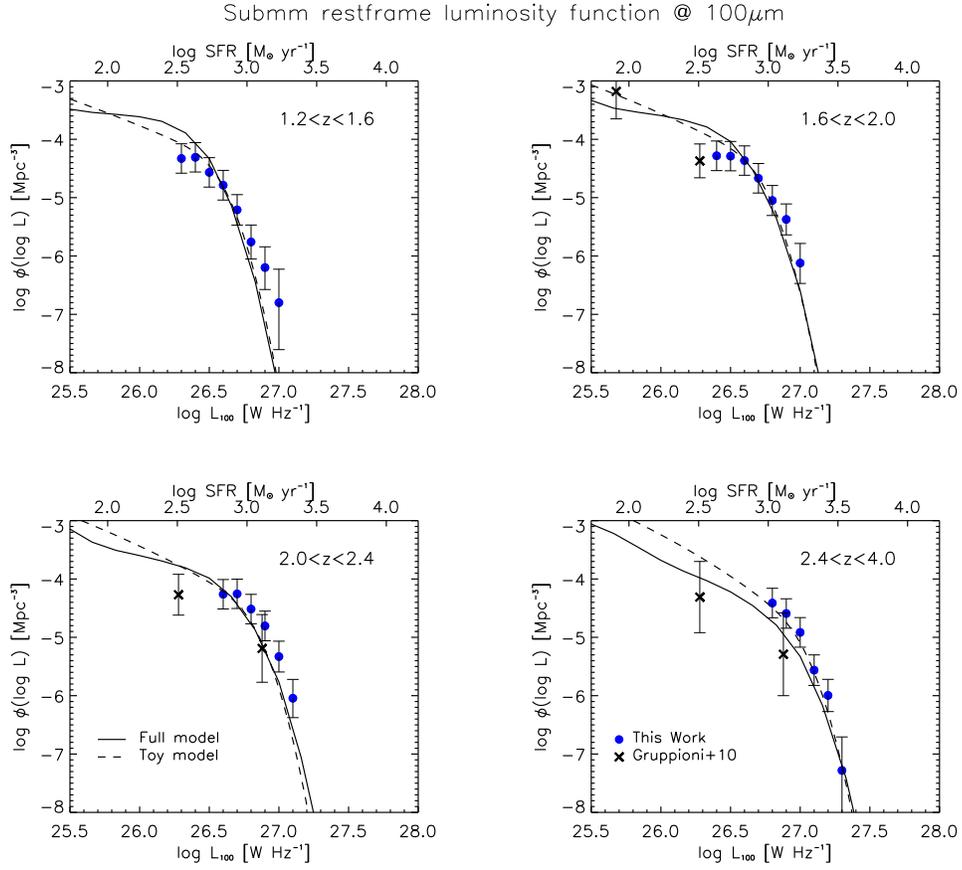}
\caption{Luminosity functions at $100\, \mu$m in four redshift intervals
compared with models described in the text (solid line: full model; dashed
line: toy model). The flattening of the LFs at the lowest luminosities may
be, at least in part, due to the overestimate of the accessible volume
yielded by the $1/V_{\rm max}$ estimator for objects near to the flux limit
(Page \& Carrera 2000). The $90\, \mu$m LFs (crosses) estimated by Gruppioni
et al. (2010) using PACS data are also shown for comparison.
}\label{fig:LF100}
\end{figure}

\clearpage
\begin{figure}
\plotone{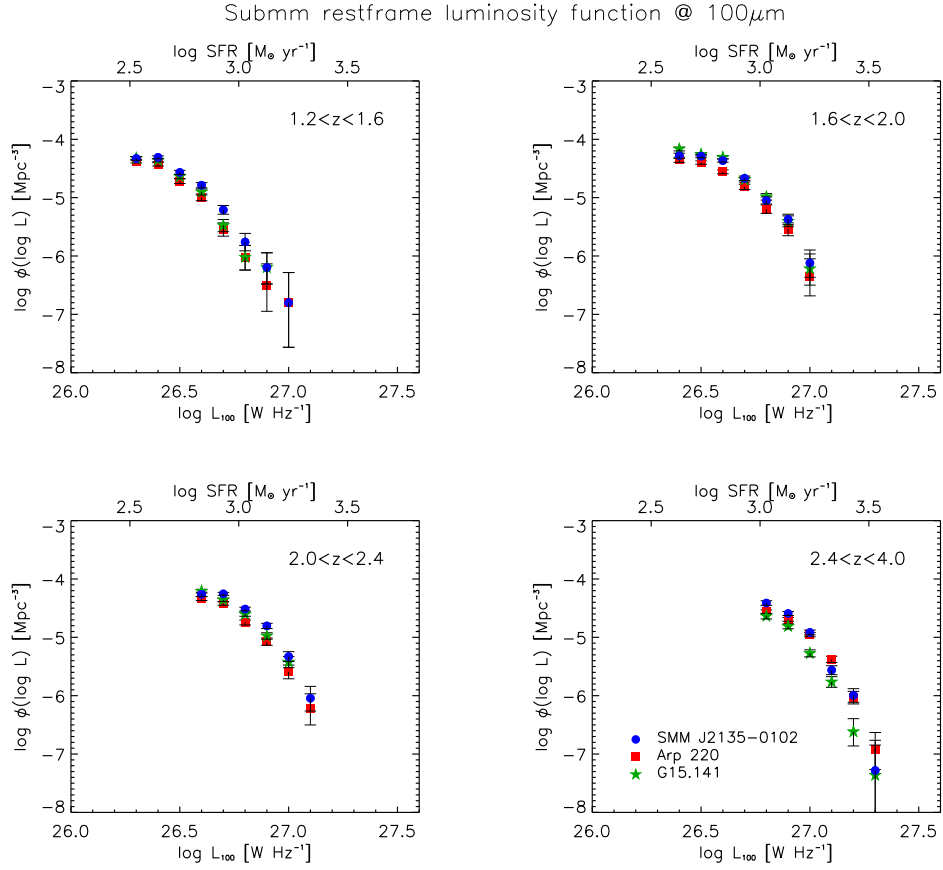}
\caption{Comparison of the luminosity functions at $100\, \mu$m computed
using different SED templates -- SMM J2135-0102 (blue circles), Arp220 (red
squares), and G15.141 (green stars) -- after correcting for the median
offsets in the redshift estimates highlighted by
Fig.~\protect\ref{fig:zspec}. To ease the comparison, only statistical
uncertainties are reported. }\label{fig:LF100sed}
\end{figure}

\clearpage
\begin{figure}
\plotone{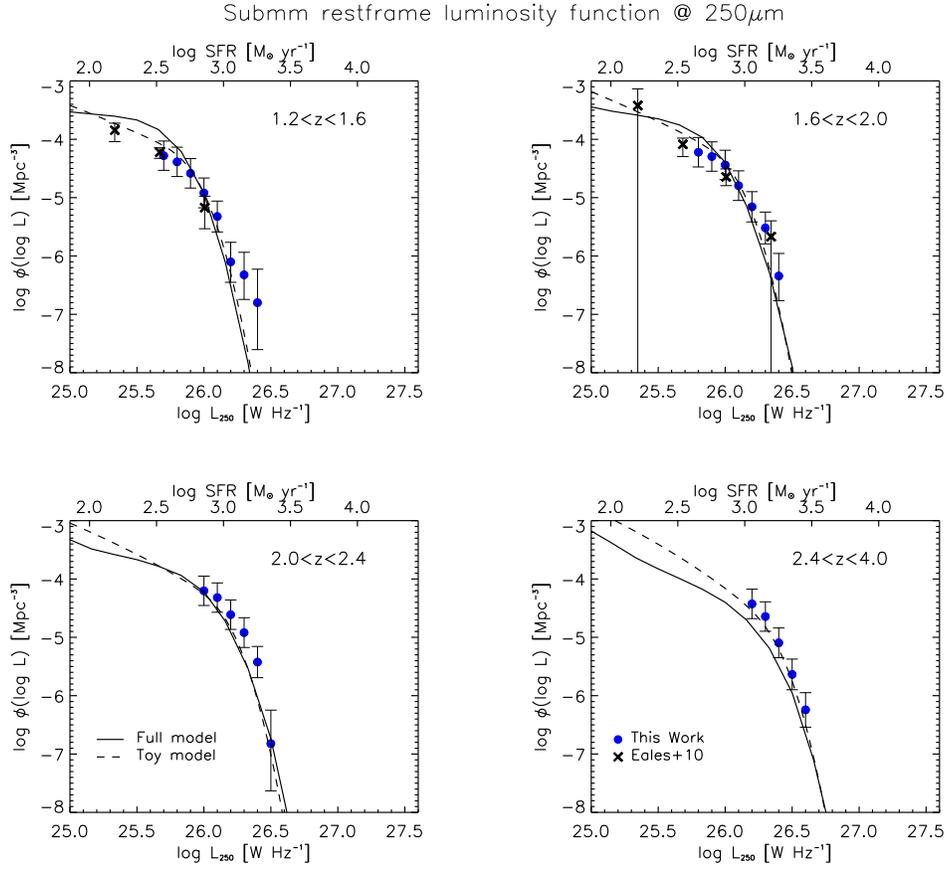}
\caption{Luminosity functions at $250\, \mu$m in four redshift intervals
based on our reference sample, compared with the predictions of the models
discussed in the text (solid line: full model; dashed line: toy model) and
with the estimates by Eales et al. (2010b), after applying the
$K-$corrections based on the SMM J2135-0102 SED, for common redshift
ranges.}\label{fig:LF250}
\end{figure}

\clearpage
\begin{figure}
\plotone{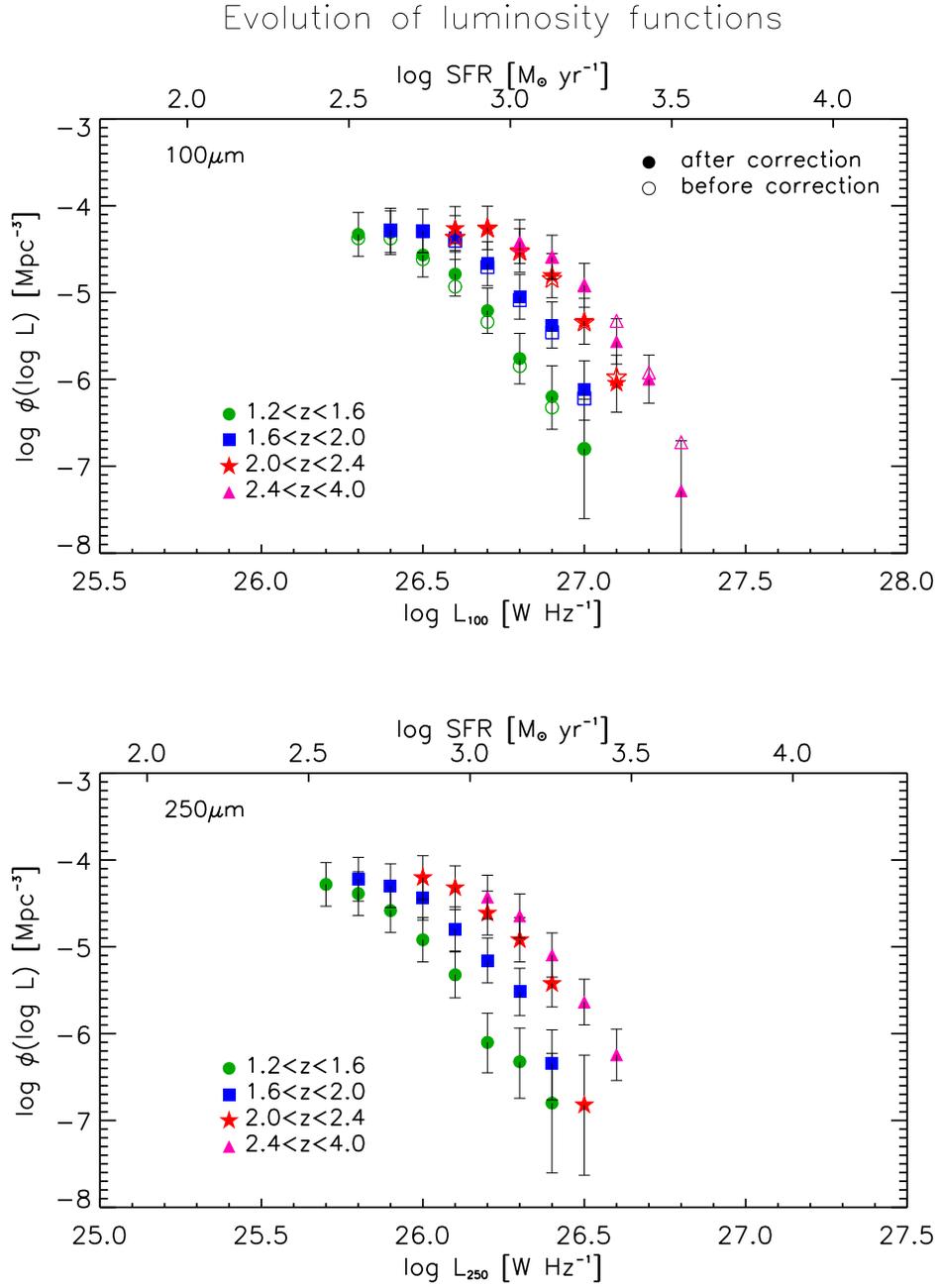} \caption{Synoptic view of the $100\, \mu$m and
$250\, \mu$m LFs for our reference sample in four redshift intervals. In the
$100\, \mu$m panel, open and filled symbols show our estimates before and
after the correction for the Eddington bias, respectively (see \S~4 for
details).}\label{fig:LF100evo}
\end{figure}

\clearpage
\begin{figure}
\plotone{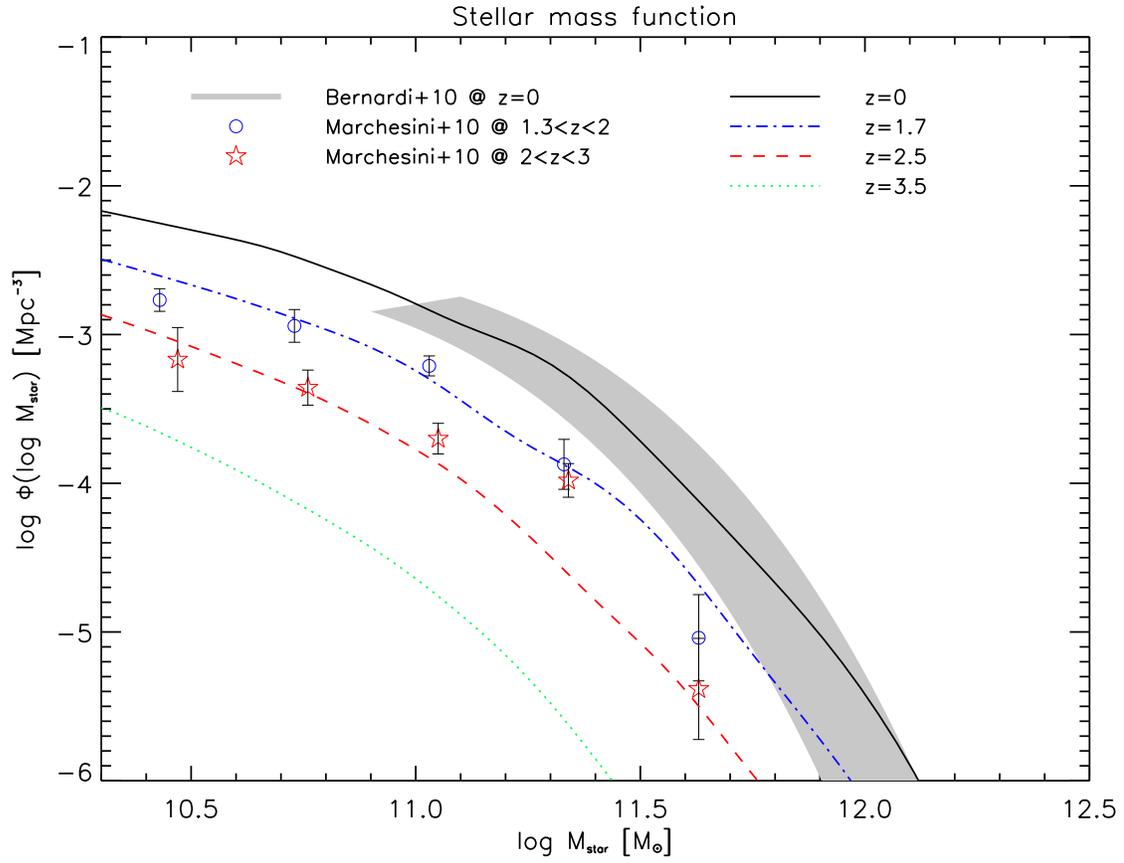} \caption{Stellar mass functions of ETGs at different
redshift predicted by the toy model compared with observational
determinations.}\label{fig:MF}
\end{figure}

\clearpage
\begin{figure}
\plotone{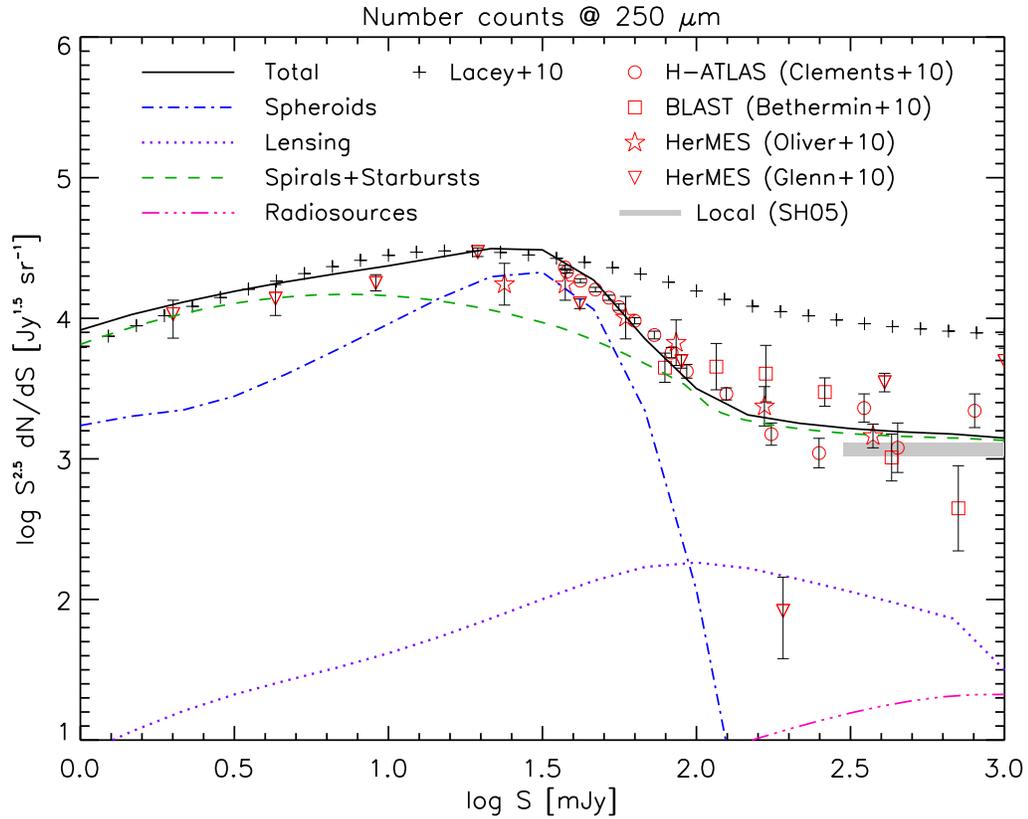} \caption{Comparison of the observed Euclidean
normalized counts at $250\,\mu$m  (Clements et al. 2010; Oliver et al. 2010;
Glenn et al. 2010; Beth\'ermin et al. 2010) with the predictions of our full
model (solid line) and of the semi-analytic model of Lacey et al. (2010;
crosses). The different lines show the contributions to the counts of the
various populations included in our full model: unlensed (blue dot-dashed
line) and strongly lensed (purple dotted line) proto-spheroidal galaxies
(host halo masses are within the range $2.5\times 10^{11}\, M_{\odot}\la
M_H\la 3\times 10^{13}\, M_{\odot}$), normal late-type plus starburst
galaxies (green dashed line), radio sources (magenta triple-dot-dashed line).
The radio source counts are from the De Zotti et al. (2005) model. The grey
shaded area shows the counts of local galaxies estimated by Serjeant \&
Harrison (2005; SH05). }\label{fig:counts250}
\end{figure}

\clearpage
\begin{figure}
\plotone{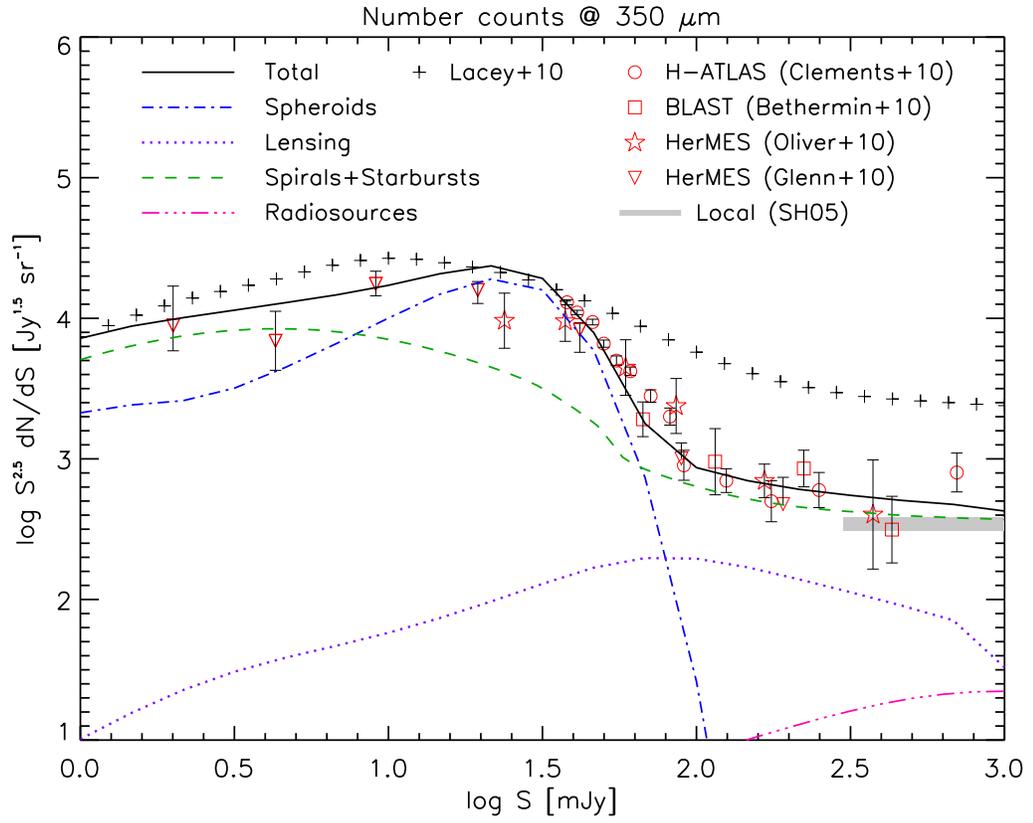} \caption{Same as in
Fig.~\protect\ref{fig:counts250} but at $350\,\mu$m.}\label{fig:counts350}
\end{figure}

\clearpage
\begin{figure}
\plotone{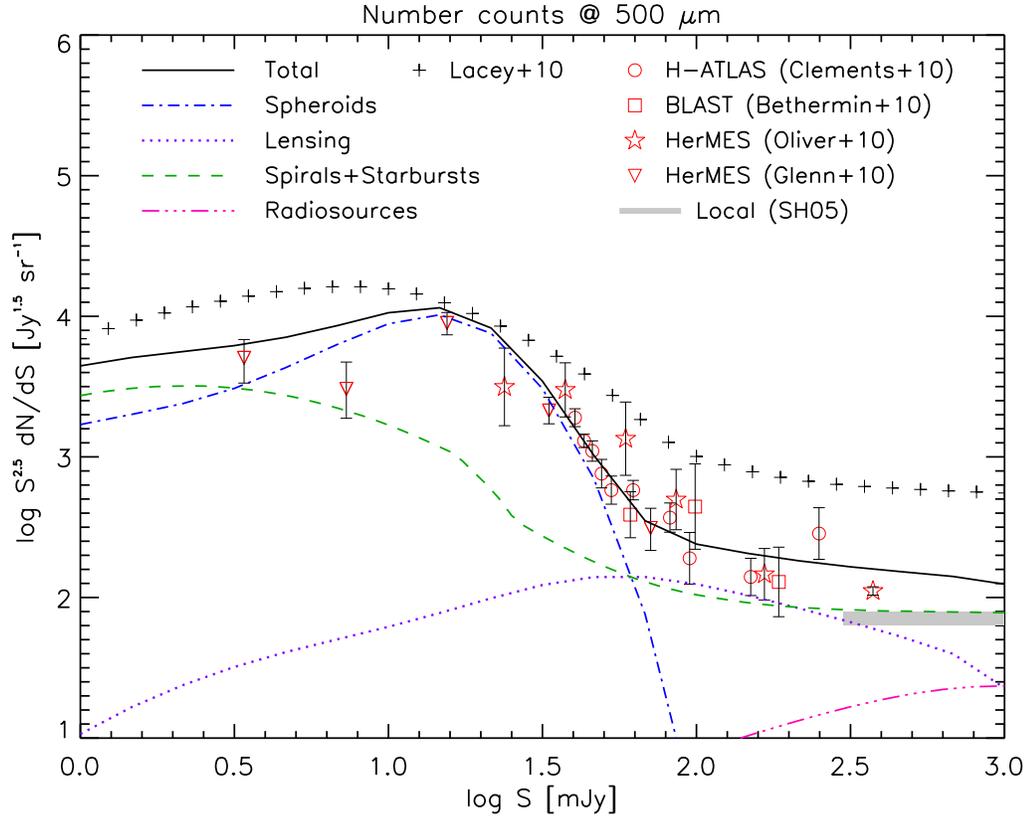} \caption{Same as in
Fig.~\protect\ref{fig:counts250} but at $500\,\mu$m. Note that the flux
densities in the Table~1 of Clements et al. (2010) must be multiplied by the
correction factors given in the same table to produce the correct flux
densities; the lowest corrected 500 micron flux density is $40$ mJy.
}\label{fig:counts500}
\end{figure}

\clearpage
\begin{figure}
\plotone{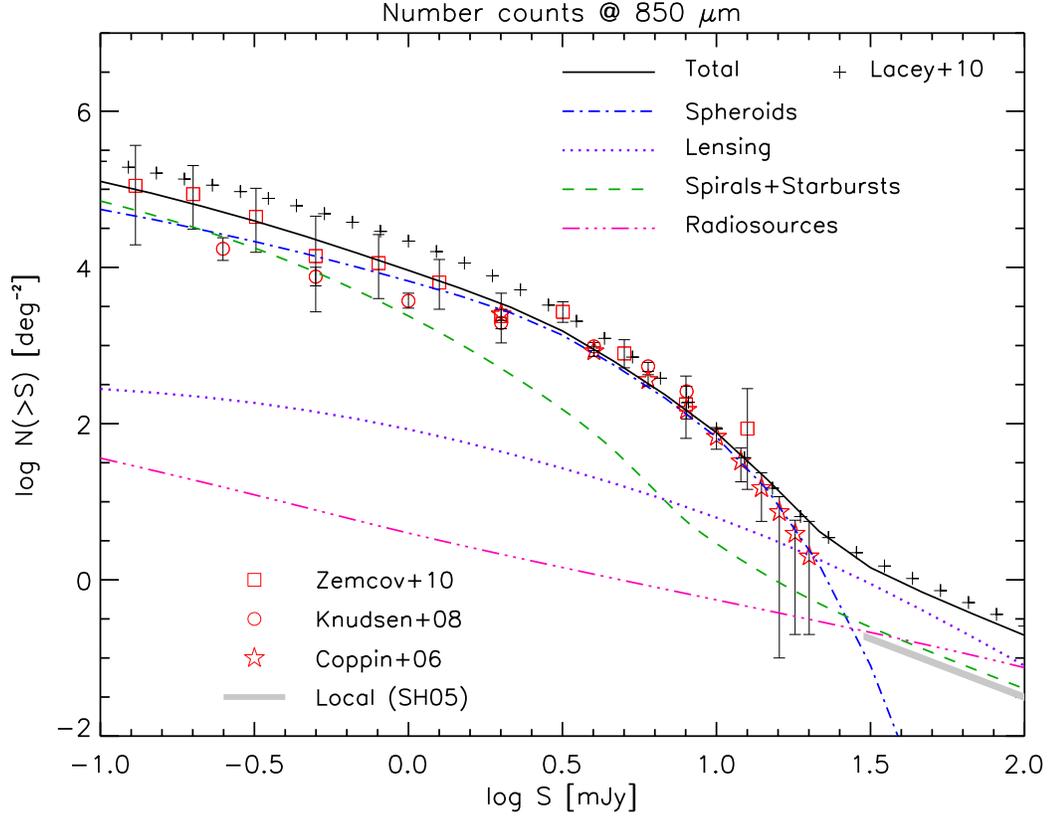} \caption{Comparison of the observed integral
counts at $850\,\mu$m  (Zemcov et al. 2010; Knudsen et al. 2008; Coppin et
al. 2006) with the predictions of our full model (solid line) and of the
semi-analytic model of Lacey et al. (2010; crosses). The lines have the same
meaning as in Fig.~\protect\ref{fig:counts250}. }\label{fig:counts850}
\end{figure}

\clearpage
\begin{figure}
\plotone{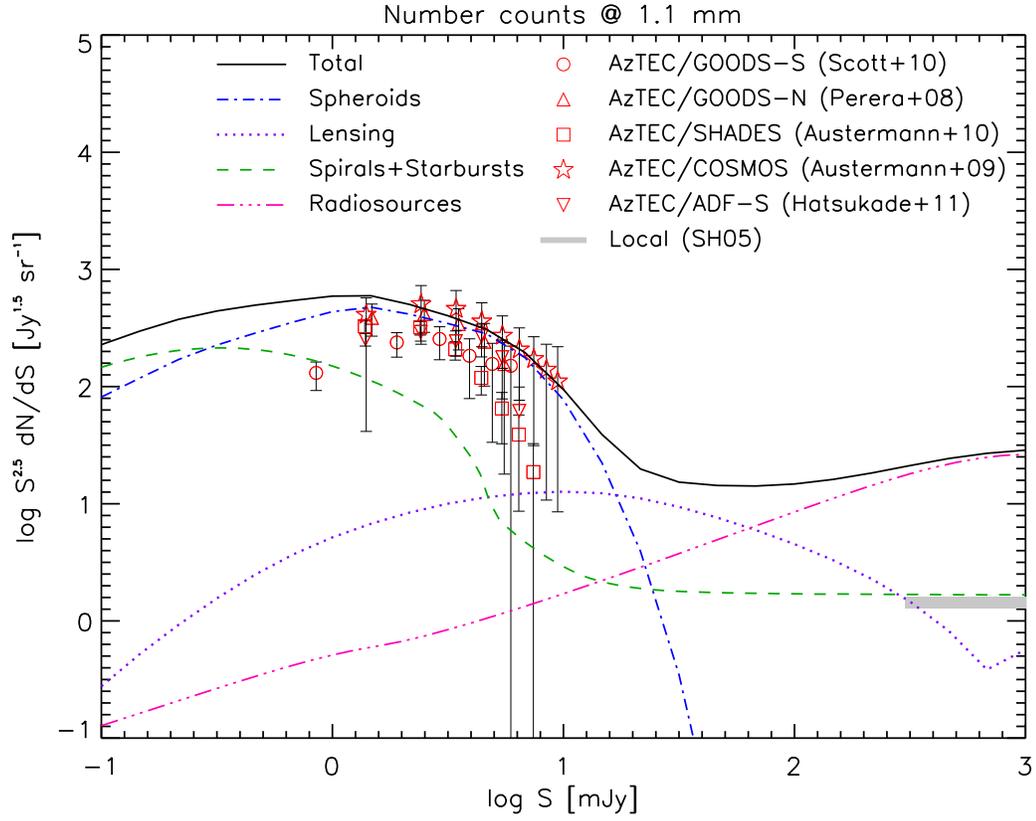} \caption{Comparison of the observed Euclidean
normalized counts at $1.1$ mm (Scott et al. 2010; Perera et al. 2008;
Austermann et al. 2009, 2010; Hatsukade et al. 2011) with the predictions of
our full model (solid line). The lines have the same meaning as in
Fig.~\protect\ref{fig:counts250}.}\label{fig:counts1100}
\end{figure}

\clearpage
\begin{figure}
\plotone{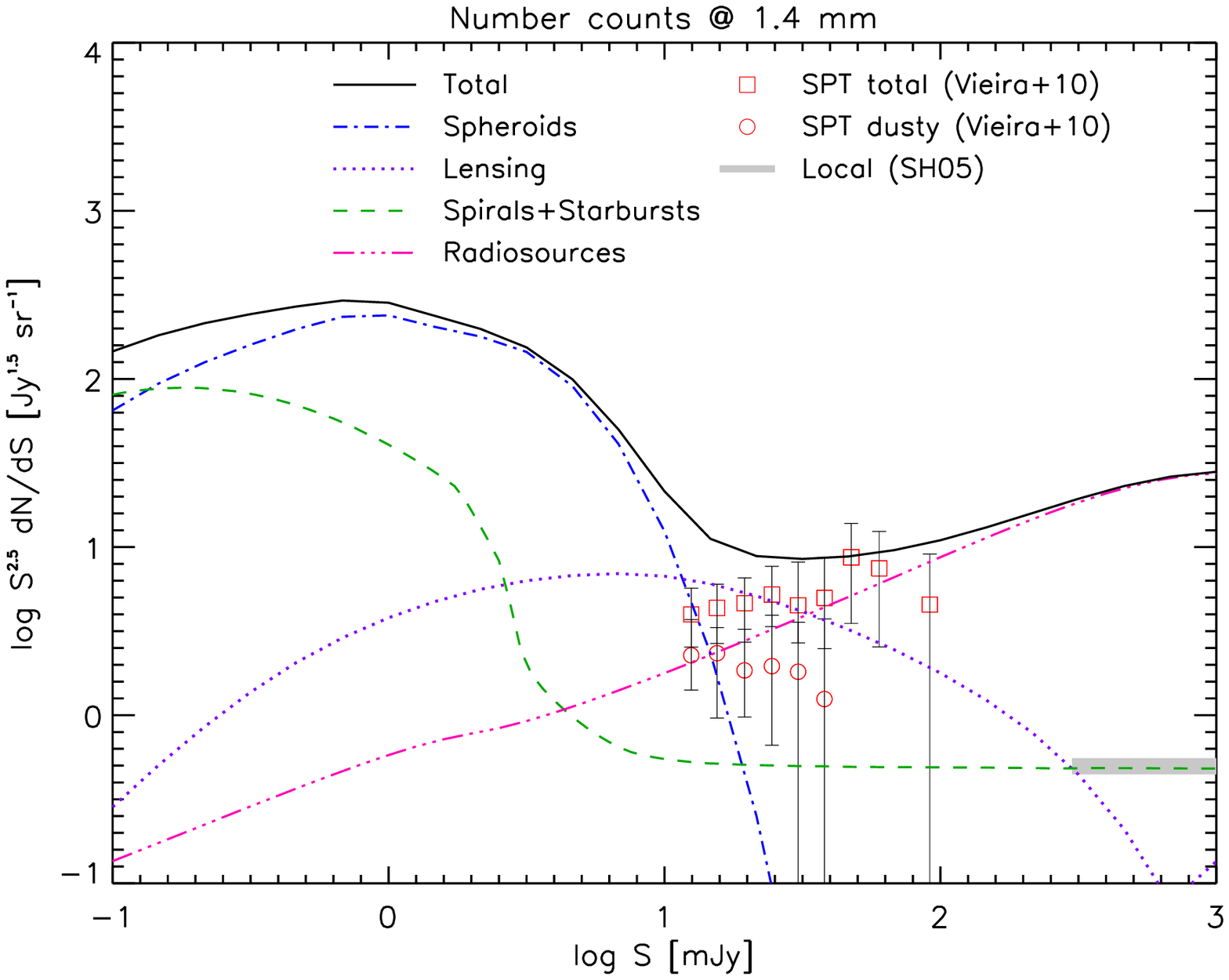} \caption{Comparison of the observed Euclidean
normalized counts at $1.4$ mm (Vieira et al. 2010) with the predictions of
our full model (solid line). The lines have the same meaning as in
Fig.~\protect\ref{fig:counts250}.}\label{fig:counts1400}
\end{figure}

\clearpage
\begin{figure}
\plotone{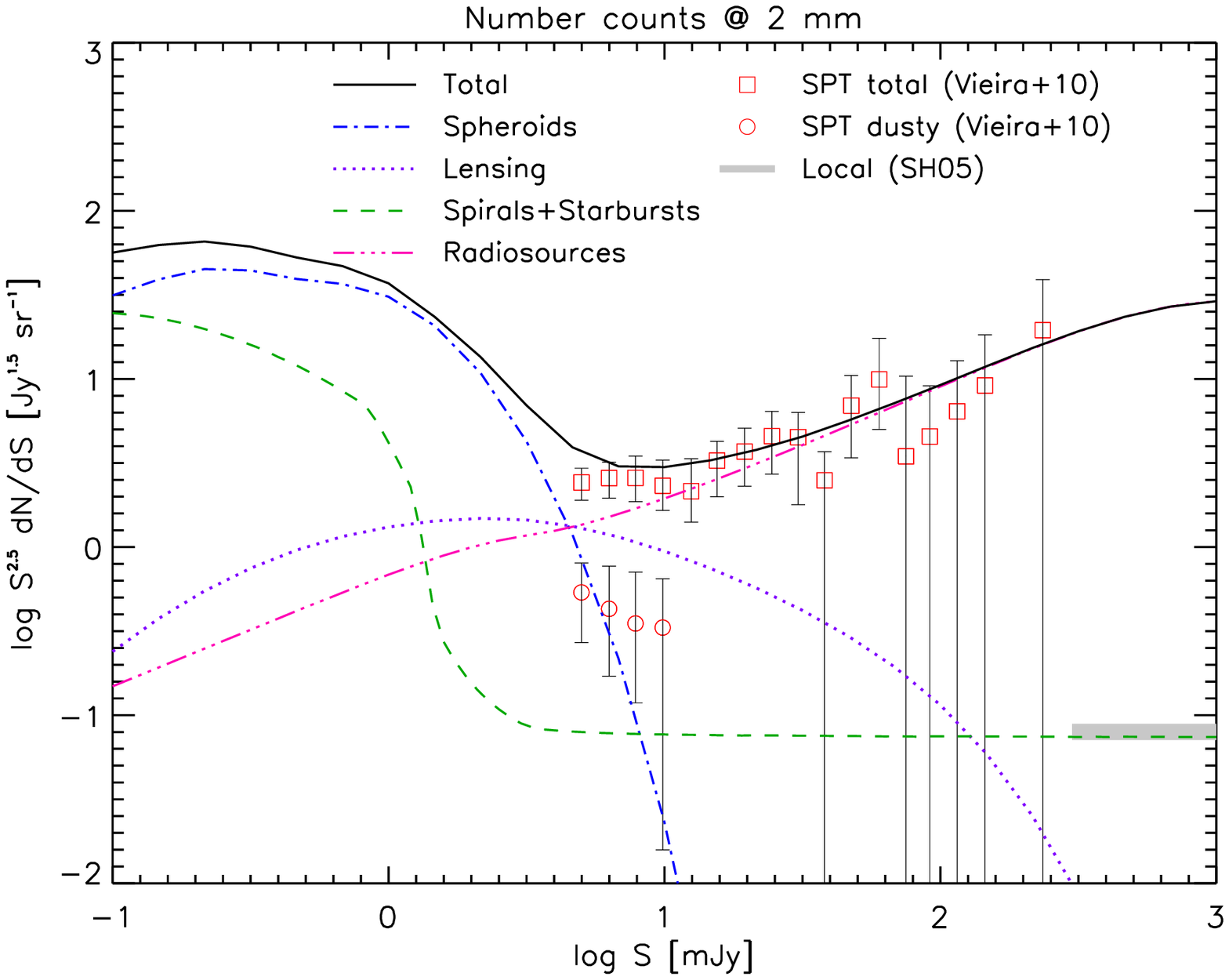} \caption{Comparison of the observed Euclidean
normalized counts at $2$ mm (Vieira et al. 2010) with the predictions of our
full model (solid line). The lines have the same meaning as in
Fig.~\protect\ref{fig:counts250}.}\label{fig:counts2000}
\end{figure}

\clearpage
\begin{figure}
\epsscale{0.6} \plotone{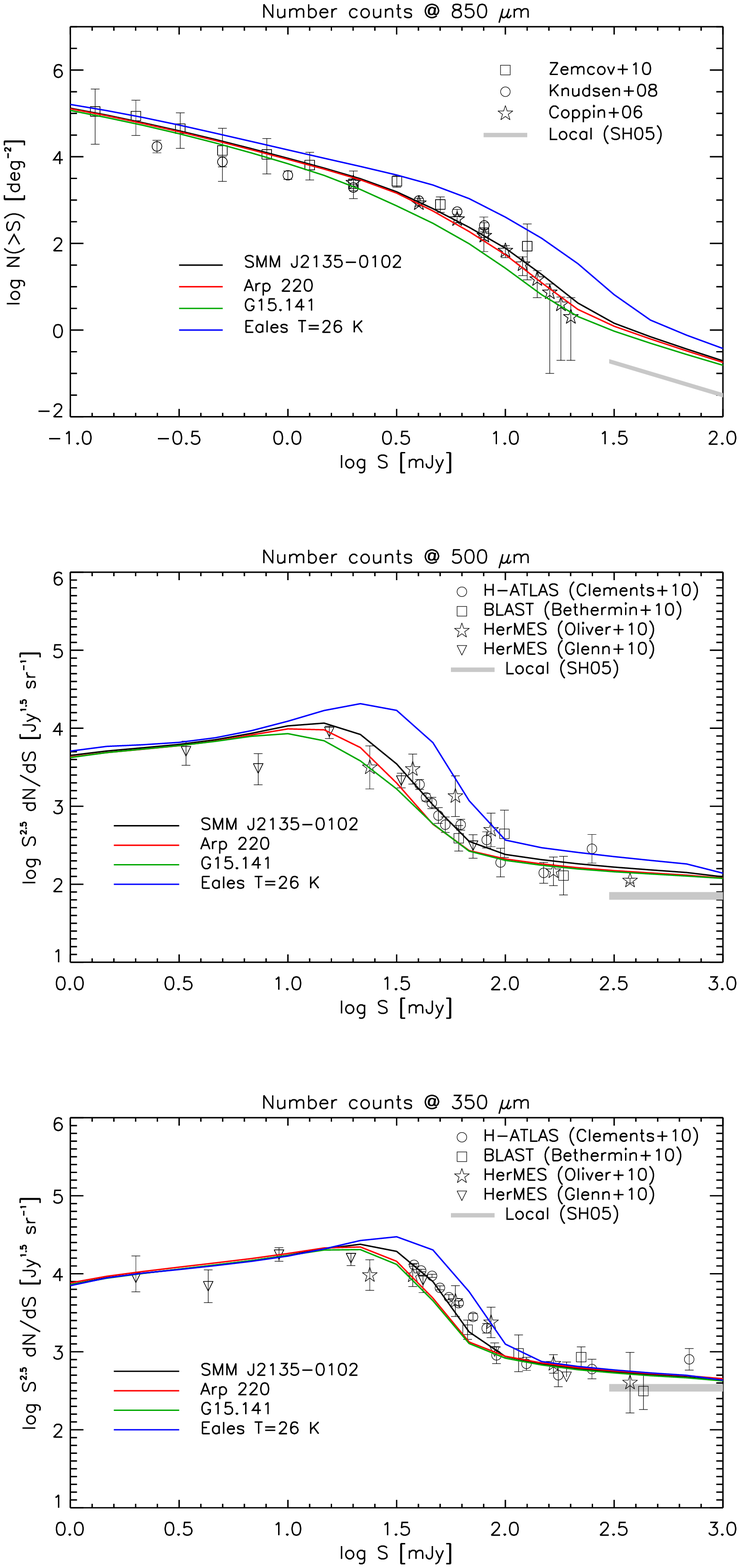} \vskip-1cm
\caption{Constraints on the SED of high-$z$ galaxies from multi-wavelength
source counts. We have determined the luminosity functions at different
redshifts $>1$ using our reference sample and the SEDs of Arp220 and G15.141,
in addition to SMM J2135-0102, and used them to compute the number counts at
several wavelengths, adding the contributions of (lower-$z$) spiral and
starburst galaxies as given by the Negrello et al. (2007) model. In all cases
we recover a good fit at $250\mu$m, essentially by construction since this is
the main selection wavelength. At longer wavelengths, however, while the
counts obtained using the SMM J2135-0102 SED are still nicely consistent with
the data, the other SEDs do not yield good fits. The counts given by the
luminosity functions estimated by Eales et al. (2010), K-corrected with their
SED, are also shown for comparison. In this case too the observed $250\mu$m
counts are accurately reproduced.}\label{fig:counts_comp}
\end{figure}

\clearpage
\begin{figure}
\epsscale{0.6}
\plotone{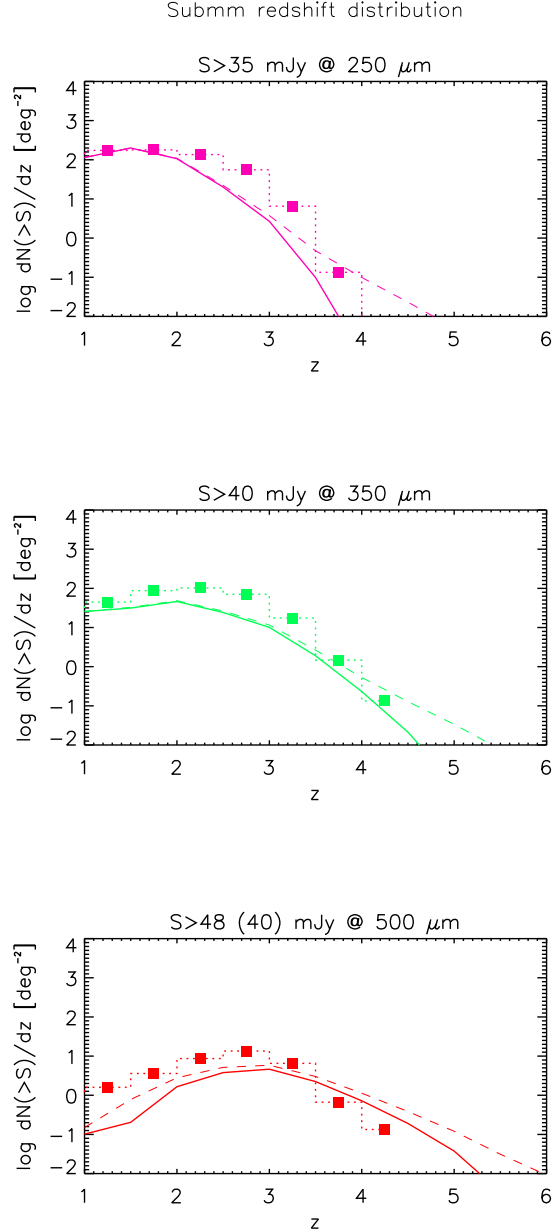} \vskip-1cm \caption{Comparison between the
estimated redshift distributions (histograms) of proto-spheroidal galaxies at
$250\mu$m ($S_{250\mu\rm m}\ge 35\,$mJy), $350\mu$m ($S_{350\mu\rm m}\ge
40\,$mJy), and $500\,\mu$m ($S_{500\mu\rm m}\ge 48\,$mJy), and the
predictions of our full model, using the same parameters as in Lapi et al.
(2006); solid lines refer to unlensed, while dashed lines also include the
contribution of lensed, proto-spheroids. Because of the flux boosting due to
confusion, the catalogued $500\,\mu$m flux densities have to be scaled down
by the factors given by Clements et al. (2010). For example, the $500\,\mu$m
flux density limit of 48 mJy for catalogued sources  corresponds to a
boosting corrected flux limit of $40\,$mJy. As discussed in the text (see
also the caption to Fig.~\protect\ref{fig:zdist}), we estimate that $\approx
55\%$ of SDP galaxies with $S_{250\mu\rm m}>35\,$mJy are at $z<1$. Such
galaxies belong to a different population comprising late-type normal and
starburst  galaxies. }\label{fig:zdist_comp}
\end{figure}

\clearpage
\begin{figure}
\epsscale{1} \plotone{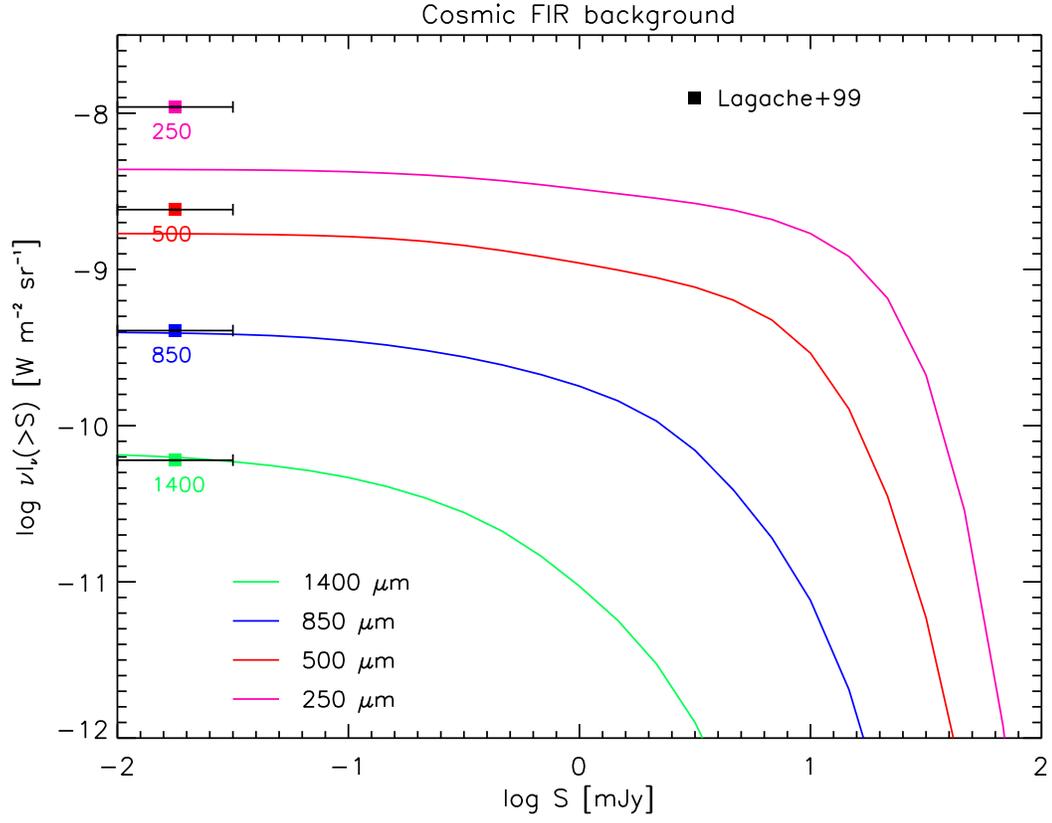} \caption{Contributions of
high-$z$ proto-spheroidal galaxies to the (sub-)mm extragalactic background,
according to our model, compared with the observational estimates of the
total background intensity (Lagache et al. 1999). ETGs account almost
entirely for the background at $\lambda\ga 850\,\mu$m. Contributions from
lower redshift ($z\la 1-1.5$) late type, normal and starburst galaxies become
increasingly important at shorter wavelengths.}\label{fig:background}
\end{figure}

\clearpage
\begin{deluxetable}{ccccccccc}
\tabletypesize{}\tablecaption{Rest-frame luminosity functions at $100\,\mu$m}
\tablewidth{0pt} \tablehead{\colhead{log $L_{100\mu\rm m}$ [W Hz$^{-1}$]} &&
\multicolumn{7}{c}{$\log \phi(\log L_{100})$ [Mpc$^{-3}$ dex$^{-1}$]}\\
\\
\cline{3-9}\\
\colhead{}&& \colhead{$1.2\le z<1.6$} && \colhead{$1.6\le z<2.0$} &&
\colhead{$2.0\le z<2.4$} && \colhead{$2.4\le z<4.0$}} \startdata
$26.3$ && $-4.329^{+0.030}_{-0.030}$ &&                            &&                            &&
\\
$26.4$ && $-4.309^{+0.025}_{-0.025}$ && $-4.283^{+0.044}_{-0.044}$ &&                            &&
\\
$26.5$ && $-4.567^{+0.034}_{-0.034}$ && $-4.289^{+0.027}_{-0.027}$ &&                            &&
\\
$26.6$ && $-4.787^{+0.044}_{-0.044}$ && $-4.365^{+0.026}_{-0.026}$ && $-4.262^{+0.036}_{-0.036}$ &&
\\
$26.7$ && $-5.209^{+0.074}_{-0.075}$ && $-4.668^{+0.037}_{-0.038}$ && $-4.255^{+0.024}_{-0.025}$ &&
\\
$26.8$ && $-5.759^{+0.146}_{-0.152}$ && $-5.049^{+0.059}_{-0.060}$ && $-4.516^{+0.031}_{-0.031}$ &&
$-4.412^{+0.039}_{-0.039}$  \\
$26.9$ && $-6.198^{+0.252}_{-0.281}$ && $-5.373^{+0.088}_{-0.090}$ && $-4.805^{+0.044}_{-0.044}$ &&
$-4.591^{+0.033}_{-0.034}$  \\
$27.0$ && $-6.800^{+0.516}_{-0.764}$ && $-6.121^{+0.223}_{-0.244}$ && $-5.330^{+0.083}_{-0.085}$ &&
$-4.915^{+0.040}_{-0.040}$  \\
$27.1$ &&                            &&                            && $-6.044^{+0.203}_{-0.219}$ &&
$-5.562^{+0.075}_{-0.076}$  \\
$27.2$ &&                            &&                            &&                            &&
$-5.995^{+0.115}_{-0.119}$  \\
$27.3$ &&                            &&                            &&                            &&
$-7.281^{+0.517}_{-0.769}$  \\
\enddata
\tablecomments{Based on our reference sample ($S_{250\mu\rm m}>35\,$mJy,
$S_{350\mu\rm m}>3\sigma$, and no $R>0.8$ optical identifications). Redshift
estimates relying on the SMM J2135-0102 SED.  Only statistical uncertainties
are quoted; an additional uncertainty of $\Delta \log \phi\approx 0.25$
related to the redshift estimate has to be summed in
quadrature.}\label{tab:lf100}
\end{deluxetable}

\clearpage
\begin{deluxetable}{ccccccccc}
\tabletypesize{}\tablecaption{Rest-frame luminosity functions at $250\,\mu$m}
\tablewidth{0pt} \tablehead{\colhead{log $L_{250\mu\rm m}$ [W Hz$^{-1}$]} &&
\multicolumn{7}{c}{$\log \phi(\log L_{250})$ [Mpc$^{-3}$ dex$^{-1}$]}\\
\\
\cline{3-9}\\
\colhead{}&& \colhead{$1.2\le z<1.6$} && \colhead{$1.6\le z<2.0$} &&
\colhead{$2.0\le z<2.4$} && \colhead{$2.4\le z<4.0$}} \startdata
$25.7$ && $-4.281^{+0.026}_{-0.026}$ &&                            &&                             &&
\\
$25.8$ && $-4.387^{+0.027}_{-0.027}$ && $-4.222^{+0.035}_{-0.035}$ &&                             &&
\\
$25.9$ && $-4.583^{+0.035}_{-0.035}$ && $-4.296^{+0.025}_{-0.025}$ &&                             &&
\\
$26.0$ && $-4.919^{+0.052}_{-0.052}$ && $-4.439^{+0.028}_{-0.028}$ &&  $-4.203^{+0.029}_{-0.029}$ &&
\\
$26.1$ && $-5.323^{+0.085}_{-0.087}$ && $-4.795^{+0.044}_{-0.044}$ &&  $-4.319^{+0.025}_{-0.025}$ &&
\\
$26.2$ && $-6.101^{+0.223}_{-0.244}$ && $-5.157^{+0.068}_{-0.068}$ &&  $-4.612^{+0.035}_{-0.035}$ &&
$-4.427^{+0.035}_{-0.035}$ \\
$26.3$ && $-6.323^{+0.294}_{-0.339}$ && $-5.519^{+0.106}_{-0.108}$ &&  $-4.919^{+0.051}_{-0.051}$ &&
$-4.643^{+0.033}_{-0.033}$ \\
$26.4$ && $-6.800^{+0.516}_{-0.764}$ && $-6.343^{+0.294}_{-0.340}$ &&  $-5.424^{+0.094}_{-0.096}$ &&
$-5.094^{+0.048}_{-0.048}$ \\
$26.5$ &&                            &&                            &&  $-6.822^{+0.517}_{-0.769}$ &&
$-5.635^{+0.080}_{-0.081}$ \\
$26.6$ &&                            &&                            &&                             &&
$-6.242^{+0.154}_{-0.161}$ \\
\enddata
\tablecomments{See note to Table~\protect\ref{tab:lf100}.}\label{tab:lf250}
\end{deluxetable}

\end{document}